\documentclass{article}
\usepackage{amsmath,amssymb}
\usepackage{supertabular}
\usepackage[a4paper, margin=1.5cm, right=1cm, top=1.8cm, bottom=2cm]{geometry} 
\usepackage{ifthen} 
\usepackage{listings} 
\usepackage[latin1]{inputenc}
\usepackage{verbatim}
\usepackage[pdftex]{color,graphicx}

\begin{document}

  \thispagestyle{empty}

\vspace*{-0.5cm}
\hspace*{-0.5cm}
\begin{minipage}{18cm}

  \noindent\includegraphics[height=2.1cm]{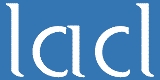}\hfill
  \includegraphics[height=2.1cm]{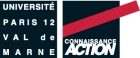}

  \vspace*{0.5cm}

  %\hrule width 20cm
  \hrule

  \vspace*{2cm}

  \begin{center}
    {\LARGE\bf A Formally Specified Type System}
  \end{center}

  \begin{center}
    {\LARGE\bf and Operational Semantics}
  \end{center}

  \begin{center}
    {\LARGE\bf for Higher-Order Procedural Variables}
  \end{center}

  \vspace*{1cm}

  \begin{center}
    {\bf Tristan Crolard and Emmanuel Polonowski}
  \end{center}

  \vspace*{5.5cm}

  \begin{center}
    {\it May\ \ 2009}\\[1ex]
    {\large\bf TR--LACL--2009--3}
  \end{center}

  \vspace*{6.5cm}

  \hrule

  {\begin{center}
      \textbf{Laboratoire d'Algorithmique, Complexit\'e et Logique (LACL)}\\
	     {\small{\bf  D\'epartement d'Informatique}\\
	       {\bf Universit\'e Paris~12 -- Val~de~Marne, Facult\'e des Science et
		 Technologie}\\
	       61, Avenue du G\'en\'eral de Gaulle, 94010 Cr\'eteil cedex, France\\
	       Tel.: (33)(1) 45 17 16 47,\ \  Fax: (33)(1) 45 17 66 01}
  \end{center}}

\end{minipage}
\nopagebreak

  \clearpage

  \thispagestyle{empty}

  \vspace*{3cm}

  \begin{center}
    Laboratory of Algorithmics, Complexity and Logic (LACL)\\
    University Paris 12 (Paris Est)\\[1ex]
    Technical Report {\bf TR--LACL--2009--3}\\[2ex]
    T.~Crolard and E.~Polonowski.\\ {\it A Formally Specified Type System and Operational Semantics \\ for Higher-Order Procedural Variables}\\
  \end{center}

  \copyright\ \  T.~Crolard and E.~Polonowski,  May 2009.

  \clearpage

\setcounter{page}{1}

\lstset{language=Ada} 

\lstset{% general command to set parameter(s) 
keywordstyle=\textsf, 
identifierstyle=\textit, % nothing happens 
stringstyle=\ttfamily, % typewriter type for strings 
showstringspaces=false,
columns=fullflexible,
emph={int,bool,true,false},
emphstyle=\textsf} 

\newcommand{\paperheader}{}
\newcommand{\papertitle}{}

\title{A Formally Specified Type System and Operational Semantics \\ for Higher-Order Procedural Variables}
\author{T. Crolard and E. Polonowski}

\date{May 14, 2009}
\maketitle

\begin{abstract}
We formally specified the type system and operational semantics of Loop$^{\omega}$  with Ott  and Isabelle/HOL proof assistant. Moreover,
both the type system and the  semantics of Loop$^{\omega}$ have been tested using Isabelle/HOL program extraction facility for inductively defined relations. In particular, the program that computes the Ackermann function type checks and behaves as expected. The main difference (apart from the choice of an Ada-like concrete syntax) with  Loop$^{\omega}$ comes from the treatment of parameter passing. Indeed, since Ott does not currently fully support $\alpha$-conversion, we rephrased the operational semantics with explicit aliasing in order to implement the {\bf out} parameter passing mode.\end{abstract}

% generated by Ott 0.10.17 from: _source-1-s.ott _source-1.ott _source-2-s.ott _source-2.ott _source-3-s.ott _source-3.ott _source-4.ott source.ott
\newcommand{\ottdrule}[4][]{{\displaystyle\frac{\begin{array}{l}#2\end{array}}{#3}\quad\ottdrulename{#4}}}
\newcommand{\ottusedrule}[1]{\[#1\]}
\newcommand{\ottpremise}[1]{ #1 \\}
\newenvironment{ottdefnblock}[3][]{ \framebox{\mbox{#2}} \quad #3 \\[0pt]}{}
\newenvironment{ottfundefnblock}[3][]{ \framebox{\mbox{#2}} \quad #3 \\[0pt]\begin{displaymath}\begin{array}{l}}{\end{array}\end{displaymath}}
\newcommand{\ottfunclause}[2]{ #1 \equiv #2 \\}
\newcommand{\ottnt}[1]{\mathit{#1}}
\newcommand{\ottmv}[1]{\mathit{#1}}
\newcommand{\ottkw}[1]{\mathbf{#1}}
\newcommand{\ottcom}[1]{\text{#1}}
\newcommand{\ottdrulename}[1]{\textsc{#1}}
\newcommand{\ottcomplu}[5]{\overline{#1}^{\,#2\in #3 #4 #5}}
\newcommand{\ottcompu}[3]{\overline{#1}^{\,#2<#3}}
\newcommand{\ottcomp}[2]{\overline{#1}^{\,#2}}
\newcommand{\ottgrammartabular}[1]{\begin{supertabular}{llcllllll}#1\end{supertabular}}
\newcommand{\ottmetavartabular}[1]{\begin{supertabular}{ll}#1\end{supertabular}}
\newcommand{\ottrulehead}[3]{$#1$ & & $#2$ & & & \multicolumn{2}{l}{#3}}
\newcommand{\ottprodline}[6]{& & $#1$ & $#2$ & $#3 #4$ & $#5$ & $#6$}
\newcommand{\ottinterrule}{\\[5.0mm]}
\newcommand{\ottafterlastrule}{\\}
\renewcommand{\ottusedrule}[1]{%
\\
\mbox{}\hfill $#1$ \hfill\mbox{}\\
\\
}
\renewcommand{\ottdrule}[3]{%
{ \displaystyle\frac{ \begin{array}{l}{#1}\end{array} }{#2} \;\; {#3} }
}
\renewcommand{\ottpremise}[1]{%
\;\;\;#1\;\;%
}
\renewcommand{\ottdrulename}[1]{%
\text{({\it #1\,})}%
}
\renewcommand{\ottkw}[1]{%
\text{\sf #1}%
}
\renewcommand{\ottnt}[1]{%
\mathit{#1}%
}
\renewcommand{\framebox}[1]{%
\mbox{$\begin{array}{|c|}\hline \begin{array}{c} #1 \end{array}\\ \hline \end{array}$}%
}
\renewenvironment{ottdefnblock}[2]{\textbf{#2} \hfill \framebox{#1}\\}{}
\renewcommand{\papertitle}{TD 1 (suite) : Expressions simples}

\renewcommand{\papertitle}{TD 2 (suite) : Commandes}

\renewcommand{\papertitle}{TD 3 (ter) : Déclarations multiples}

\renewcommand{\papertitle}{TD 5 : Procédures }
\renewcommand{\ottcomp}[2]{|#1|}

\renewcommand{\papertitle}{Type system and SOS for {\sc Loop}$^\omega$}
\renewcommand{\ottcomp}[2]{|#1|}

\newcommand{\ottmetavars}{
\ottmetavartabular{
 $ \ottmv{index} ,\, \ottmv{i} ,\, \ottmv{j} ,\, \ottmv{l} ,\, \ottmv{n} $ & \multicolumn{1}{l}{\ottcom{indices}} \\
 $ \mathit{ident} ,\, \mathit{x} ,\, \mathit{y} ,\, \mathit{z} ,\, \mathit{p} ,\, \mathit{f} $ & \multicolumn{1}{l}{\ottcom{idents}} \\
 $ \ottmv{number} ,\, \ottmv{q} $ &  \\
}}

\newcommand{\ottterminals}{
\ottrulehead{\ottnt{terminals}}{::=}{}\\ 
\ottprodline{|}{ \longrightarrow }{}{}{}{}\\ 
\ottprodline{|}{ \rightarrow }{}{}{}{}\\ 
\ottprodline{|}{ \Rightarrow }{}{}{}{}\\ 
\ottprodline{|}{ \leftarrow }{}{}{}{}\\ 
\ottprodline{|}{ \mapsto }{}{}{}{}\\ 
\ottprodline{|}{ \leadsto }{}{}{}{}\\ 
\ottprodline{|}{ \vdash }{}{}{}{}\\ 
\ottprodline{|}{ \varnothing }{}{}{}{}\\ 
\ottprodline{|}{ \times }{}{}{}{}\\ 
\ottprodline{|}{ \neq }{}{}{}{}\\ 
\ottprodline{|}{ := }{}{}{}{}\\ 
\ottprodline{|}{ \langle }{}{}{}{}\\ 
\ottprodline{|}{ \rangle }{}{}{}{}\\ 
\ottprodline{|}{ \sim }{}{}{}{}\\ 
\ottprodline{|}{ \not\in }{}{}{}{}\\ 
\ottprodline{|}{ \triangleright }{}{}{}{}}

\newcommand{\ottmode}{
\ottrulehead{\ottnt{mode}  ,\ \ottnt{m}}{::=}{\ottcom{modes:}}\\ 
\ottprodline{|}{} {\textsf{S}}{}{}{}\\ 
\ottprodline{|}{\ottkw{in}}{}{}{}{}\\ 
\ottprodline{|}{\ottkw{out}}{}{}{}{}\\ 
\ottprodline{|}{\ottkw{in} \, \ottkw{out}}{}{}{}{}}

\newcommand{\ottinteger}{
\ottrulehead{\ottnt{integer}  ,\ \ottnt{k}}{::=}{}\\ 
\ottprodline{|}{\ottmv{q}}{}{}{}{}\\ 
\ottprodline{|}{\ottkw{\{} \, \ottnt{k_{{\mathrm{1}}}} \, + \, \ottnt{k_{{\mathrm{2}}}} \, \ottkw{\}}}{}{}{}{}\\ 
\ottprodline{|}{\ottkw{\{} \, \ottnt{k_{{\mathrm{1}}}} \, - \, \ottnt{k_{{\mathrm{2}}}} \, \ottkw{\}}}{}{}{}{}\\ 
\ottprodline{|}{\ottkw{\{} \, \ottnt{k_{{\mathrm{1}}}} \, \times \, \ottnt{k_{{\mathrm{2}}}} \, \ottkw{\}}}{}{}{}{}}

\newcommand{\ottboolean}{
\ottrulehead{\ottnt{boolean}  ,\ \ottnt{b}}{::=}{}\\ 
\ottprodline{|}{\ottkw{true}}{}{}{}{}\\ 
\ottprodline{|}{\ottkw{false}}{}{}{}{}\\ 
\ottprodline{|}{\ottkw{\{} \, \ottnt{b_{{\mathrm{1}}}} \, \ottkw{and} \, \ottnt{b_{{\mathrm{2}}}} \, \ottkw{\}}}{}{}{}{}\\ 
\ottprodline{|}{\ottkw{\{} \, \ottnt{b_{{\mathrm{1}}}} \, \ottkw{or} \, \ottnt{b_{{\mathrm{2}}}} \, \ottkw{\}}}{}{}{}{}\\ 
\ottprodline{|}{\ottkw{\{} \, \ottkw{not} \, \ottnt{b} \, \ottkw{\}}}{}{}{}{}\\ 
\ottprodline{|}{\ottkw{\{} \, \ottnt{k_{{\mathrm{1}}}} \, = \, \ottnt{k_{{\mathrm{2}}}} \, \ottkw{\}}}{}{}{}{}\\ 
\ottprodline{|}{ \ottnt{k_{{\mathrm{1}}}}  >  \ottnt{k_{{\mathrm{2}}}} }{}{}{}{}\\ 
\ottprodline{|}{ \ottnt{k_{{\mathrm{1}}}}  <  \ottnt{k_{{\mathrm{2}}}} }{}{}{}{}}

\newcommand{\ottexp}{
\ottrulehead{\ottnt{exp}  ,\ \ottnt{e}}{::=}{\ottcom{terms:}}\\ 
\ottprodline{|}{\mathit{x}}{}{}{}{\ottcom{var}}\\ 
\ottprodline{|}{\ottnt{v}}{}{}{}{\ottcom{value}}\\ 
\ottprodline{|}{\ottnt{e_{{\mathrm{1}}}} \, + \, \ottnt{e_{{\mathrm{2}}}}}{}{}{}{\ottcom{addition}}\\ 
\ottprodline{|}{\ottnt{e_{{\mathrm{1}}}} \, - \, \ottnt{e_{{\mathrm{2}}}}}{}{}{}{\ottcom{subtraction}}\\ 
\ottprodline{|}{\ottnt{e_{{\mathrm{1}}}} \, \times \, \ottnt{e_{{\mathrm{2}}}}}{}{}{}{\ottcom{multiplication}}\\ 
\ottprodline{|}{\ottnt{e_{{\mathrm{1}}}} \, = \, \ottnt{e_{{\mathrm{2}}}}}{}{}{}{\ottcom{equality}}\\ 
\ottprodline{|}{ \ottnt{e_{{\mathrm{1}}}}  >  \ottnt{e_{{\mathrm{2}}}} }{}{}{}{\ottcom{greater}}\\ 
\ottprodline{|}{ \ottnt{e_{{\mathrm{1}}}}  <  \ottnt{e_{{\mathrm{2}}}} }{}{}{}{\ottcom{less}}\\ 
\ottprodline{|}{\ottnt{e_{{\mathrm{1}}}} \, \ottkw{and} \, \ottnt{e_{{\mathrm{2}}}}}{}{}{}{\ottcom{conjunction}}\\ 
\ottprodline{|}{\ottnt{e_{{\mathrm{1}}}} \, \ottkw{or} \, \ottnt{e_{{\mathrm{2}}}}}{}{}{}{\ottcom{disjunction}}\\ 
\ottprodline{|}{\ottkw{not} \, \ottnt{e}}{}{}{}{\ottcom{negation}}\\ 
\ottprodline{|}{( \, \ottnt{e} \, )} {\textsf{S}}{}{}{\ottcom{parentheses}}}

\newcommand{\ottstore}{
\ottrulehead{\ottnt{store}  ,\ \mu}{::=}{\ottcom{store}}\\ 
\ottprodline{|}{\ottkw{\mbox{?}}} {\textsf{S}}{}{}{}\\ 
\ottprodline{|}{\ottkw{[]}}{}{}{}{}\\ 
\ottprodline{|}{( \, \mu \, , \, \mathit{x} \, \leftarrow \, \ottnt{v} \, )}{}{}{}{}\\ 
\ottprodline{|}{[ \, \mathit{z_{{\mathrm{1}}}} \, \leftarrow \, \ottnt{v_{{\mathrm{1}}}} \, , \, ... \, , \, \mathit{z_{\ottmv{n}}} \, \leftarrow \, \ottnt{v_{\ottmv{n}}} \, ]}{}{}{}{}}

\newcommand{\otttrace}{
\ottrulehead{\ottnt{trace}  ,\ \ottnt{tr}}{::=}{}\\ 
\ottprodline{|}{\ottkw{\mbox{?}}} {\textsf{S}}{}{}{}\\ 
\ottprodline{|}{[ \, \langle \, \ottnt{c_{{\mathrm{1}}}} \, , \, \mu_{{\mathrm{1}}} \, \rangle \, .. \, \langle \, \ottnt{c_{\ottmv{n}}} \, , \, \mu_{\ottmv{n}} \, \rangle \, ]}{}{}{}{}}

\newcommand{\ottformula}{
\ottrulehead{\ottnt{formula}}{::=}{}\\ 
\ottprodline{|}{\ottnt{formula_{{\mathrm{1}}}} \quad .. \quad \ottnt{formula_{\ottmv{n}}}}{}{}{}{}\\ 
\ottprodline{|}{\ottnt{judgement}}{}{}{}{}\\ 
\ottprodline{|}{\mathit{x} \, = \, \mathit{x'}}{}{}{}{}\\ 
\ottprodline{|}{\mathit{x} \, \neq \, \mathit{x'}}{}{}{}{}\\ 
\ottprodline{|}{\delta \, = \, \delta'}{}{}{}{}\\ 
\ottprodline{|}{\delta \, \neq \, \delta'}{}{}{}{}\\ 
\ottprodline{|}{\ottnt{m} \, = \, \ottnt{m'}}{}{}{}{}\\ 
\ottprodline{|}{\ottnt{m} \, \neq \, \ottnt{m'}}{}{}{}{}\\ 
\ottprodline{|}{ \ottnt{k}  >  \ottnt{k'} }{}{}{}{}\\ 
\ottprodline{|}{ \ottnt{k}  \leq  \ottnt{k'} }{}{}{}{}}

\newcommand{\ottenv}{
\ottrulehead{\ottnt{env}  ,\ \Gamma}{::=}{\ottcom{contexts:}}\\ 
\ottprodline{|}{\ottkw{\{\}}}{}{}{}{\ottcom{empty context}}\\ 
\ottprodline{|}{\ottkw{\{} \, \mathit{x_{{\mathrm{1}}}} \, \delta_{{\mathrm{1}}} \, , \, ... \, , \, \mathit{x_{\ottmv{n}}} \, \delta_{\ottmv{n}} \, \ottkw{\}}}{}{}{}{\ottcom{explicit context}}\\ 
\ottprodline{|}{\Gamma \, , \, \mathit{x} \, \delta}{}{}{}{\ottcom{ident declaration}}\\ 
\ottprodline{|}{ \Gamma } {\textsf{S}}{}{}{\ottcom{parentheses}}\\ 
\ottprodline{|}{\Gamma \, , \, \mathit{x_{{\mathrm{1}}}} \, \delta_{{\mathrm{1}}} \, , \, .... \, , \, \mathit{x_{\ottmv{n}}} \, \delta_{\ottmv{n}}}{}{}{}{\ottcom{idents declaration}}\\ 
\ottprodline{|}{\Gamma \, , \, \delta}{}{}{}{\ottcom{anonymous declaration}}}

\newcommand{\ottcmd}{
\ottrulehead{\ottnt{cmd}  ,\ \ottnt{c}}{::=}{\ottcom{commands:}}\\ 
\ottprodline{|}{\ottkw{null}}{}{}{}{\ottcom{null}}\\ 
\ottprodline{|}{\mathit{x} \, := \, \ottnt{e}}{}{}{}{\ottcom{assignment}}\\ 
\ottprodline{|}{\ottnt{c_{{\mathrm{1}}}} \, ; \, \ottnt{c_{{\mathrm{2}}}}}{}{}{}{\ottcom{sequence}}\\ 
\ottprodline{|}{\ottkw{if} \, \ottnt{e} \, \ottkw{then} \, \ottnt{c_{{\mathrm{1}}}} \, ; \, \ottkw{else} \, \ottnt{c_{{\mathrm{2}}}} \, ; \, \ottkw{end} \, \ottkw{if}}{}{}{}{\ottcom{conditional}}\\ 
\ottprodline{|}{\ottkw{while} \, \ottnt{e} \, \ottkw{loop} \, \ottnt{c} \, ; \, \ottkw{end} \, \ottkw{loop}}{}{}{}{\ottcom{while loop}}\\ 
\ottprodline{|}{( \, \ottnt{c} \, )} {\textsf{S}}{}{}{}\\ 
\ottprodline{|}{\ottkw{\mbox{?}}} {\textsf{S}}{}{}{}\\ 
\ottprodline{|}{\ottkw{declare} \, \ottnt{d}}{}{}{}{\ottcom{declaration}}\\ 
\ottprodline{|}{\ottkw{for} \, \mathit{x} \, \ottkw{in} \, \ottnt{e} \, . \, . \, \ottnt{e'} \, \ottkw{loop} \, \ottnt{c} \, ; \, \ottkw{end} \, \ottkw{loop}}{}{\textsf{bind}\; \mathit{x}\; \textsf{in}\; \ottnt{c}}{}{\ottcom{for loop}}\\ 
\ottprodline{|}{\ottnt{e} \, ( \, \ottnt{e_{{\mathrm{1}}}} \, , \, .. \, , \, \ottnt{e_{\ottmv{n}}} \, )}{}{}{}{\ottcom{Procedure call}}}

\newcommand{\ottva}{
\ottrulehead{\ottnt{va}  ,\ \ottnt{v}}{::=}{\ottcom{constants:}}\\ 
\ottprodline{|}{\ottkw{\mbox{?}}} {\textsf{S}}{}{}{}\\ 
\ottprodline{|}{\ottnt{k}}{}{}{}{\ottcom{integer constant}}\\ 
\ottprodline{|}{\ottnt{b}}{}{}{}{\ottcom{boolean true}}\\ 
\ottprodline{|}{\ottkw{proc} \, ( \, \mathit{x_{{\mathrm{1}}}} \, : \, \ottnt{m_{{\mathrm{1}}}} \, \tau_{{\mathrm{1}}} \, ; \, .. \, ; \, \mathit{x_{\ottmv{n}}} \, : \, \ottnt{m_{\ottmv{n}}} \, \tau_{\ottmv{n}} \, ) \, \ottkw{is} \, \ottnt{d}}{}{\textsf{bind}\; \mathit{x_{{\mathrm{1}}}}..\mathit{x_{\ottmv{n}}}\; \textsf{in}\; \ottnt{d}}{}{}}

\newcommand{\ottty}{
\ottrulehead{\ottnt{ty}  ,\ \tau}{::=}{\ottcom{types:}}\\ 
\ottprodline{|}{\ottkw{int}}{}{}{}{}\\ 
\ottprodline{|}{\ottkw{bool}}{}{}{}{}\\ 
\ottprodline{|}{\ottkw{proc} \, ( \, \ottnt{m_{{\mathrm{1}}}} \, \tau_{{\mathrm{1}}} \, , \, .. \, , \, \ottnt{m_{\ottmv{n}}} \, \tau_{\ottmv{n}} \, )}{}{}{}{\ottcom{Procedure}}\\ 
\ottprodline{|}{\ottkw{void}}{}{}{}{\ottcom{void}}}

\newcommand{\ottdf}{
\ottrulehead{\delta}{::=}{}\\ 
\ottprodline{|}{: \, \ottnt{m} \, \tau}{}{}{}{}\\ 
\ottprodline{|}{ \hookrightarrow  \tau }{}{}{}{}}

\newcommand{\ottdcl}{
\ottrulehead{\ottnt{dcl}  ,\ \ottnt{d}}{::=}{}\\ 
\ottprodline{|}{\ottnt{d} \, [ \, \ottnt{v} \, / \, \mathit{x} \, ]} {\textsf{M}}{}{}{}\\ 
\ottprodline{|}{\ottkw{begin} \, \ottkw{end}}{}{}{}{\ottcom{Empty}}\\ 
\ottprodline{|}{\ottkw{begin} \, \ottnt{c} \, ; \, \ottkw{end}}{}{}{}{\ottcom{Block}}\\ 
\ottprodline{|}{\mathit{x} \, : \, \tau \, ; \, \ottnt{d}}{}{\textsf{bind}\; \mathit{x}\; \textsf{in}\; \ottnt{d}}{}{\ottcom{Uninit. variable}}\\ 
\ottprodline{|}{\mathit{x} \, : \, \tau \, := \, \ottnt{e} \, ; \, \ottnt{d}}{}{\textsf{bind}\; \mathit{x}\; \textsf{in}\; \ottnt{d}}{}{\ottcom{Init. variable}}\\ 
\ottprodline{|}{\mathit{x} \, : \, \ottkw{constant} \, \tau \, := \, \ottnt{e} \, ; \, \ottnt{d}}{}{\textsf{bind}\; \mathit{x}\; \textsf{in}\; \ottnt{d}}{}{\ottcom{Constant}}\\ 
\ottprodline{|}{\ottkw{procedure} \, \mathit{p} \, ( \, \mathit{x_{{\mathrm{1}}}} \, : \, \ottnt{m_{{\mathrm{1}}}} \, \tau_{{\mathrm{1}}} \, ; \, .. \, ; \, \mathit{x_{\ottmv{n}}} \, : \, \ottnt{m_{\ottmv{n}}} \, \tau_{\ottmv{n}} \, ) \, \ottkw{is} \, \ottnt{d_{{\mathrm{1}}}} \, ; \, \ottnt{d_{{\mathrm{2}}}}}{}{\textsf{bind}\; \mathit{x_{{\mathrm{1}}}}..\mathit{x_{\ottmv{n}}}\; \textsf{in}\; \ottnt{d_{{\mathrm{1}}}}}{}{\ottcom{Proc}}\\ 
\ottprodline{|}{[ \, \mathit{x_{{\mathrm{1}}}} \, : \, \ottnt{m_{{\mathrm{1}}}} \, \tau_{{\mathrm{1}}} \, = \, \ottnt{e_{{\mathrm{1}}}} \, , \, .. \, , \, \mathit{x_{\ottmv{n}}} \, : \, \ottnt{m_{\ottmv{n}}} \, \tau_{\ottmv{n}} \, = \, \ottnt{e_{\ottmv{n}}} \, ] \, \ottnt{d}}{}{\textsf{bind}\; \mathit{x_{{\mathrm{1}}}}..\mathit{x_{\ottmv{n}}}\; \textsf{in}\; \ottnt{d}}{}{\ottcom{Aliases}}\\ 
\ottprodline{|}{( \, \mathit{x} \, : \, \ottnt{m} \, \tau \, = \, \ottnt{e} \, ) \, \ottnt{d}}{}{\textsf{bind}\; \mathit{x}\; \textsf{in}\; \ottnt{d}}{}{\ottcom{Alias}}}

\newcommand{\ottvalue}{
\ottrulehead{\ottnt{value}}{::=}{}\\ 
\ottprodline{|}{\ottnt{v}}{}{}{}{\ottcom{value}}}

\newcommand{\ottevalXXexp}{
\ottrulehead{\ottnt{eval\_exp}}{::=}{}\\ 
\ottprodline{|}{\mu \, ( \, \mathit{x} \, ) \, = \, \ottnt{v}}{}{}{}{\ottcom{Fetch}}\\ 
\ottprodline{|}{ \ottnt{e}  =_ \mu   \ottnt{v} }{}{}{}{\ottcom{Expression evaluation}}}

\newcommand{\otttyping}{
\ottrulehead{\ottnt{typing}}{::=}{}\\ 
\ottprodline{|}{ \mathit{ \mathit{x} } \;  \delta  \; \in \;  \Gamma }{}{}{}{\ottcom{Lookup}}\\ 
\ottprodline{|}{\Gamma \, \vdash \, \ottnt{e} \, : \, \tau}{}{}{}{\ottcom{Expression typing}}\\ 
\ottprodline{|}{ \delta  \; \in \;  \Gamma }{}{}{}{\ottcom{LookupD}}\\ 
\ottprodline{|}{\Gamma \, \vdash \, \ottnt{e} \, \sim \, \ottnt{m} \, \tau}{}{}{}{\ottcom{Match}}\\ 
\ottprodline{|}{\Gamma \, \vdash \, ( \, \ottnt{e_{{\mathrm{1}}}} \, , \, .. \, , \, \ottnt{e_{\ottmv{l}}} \, ) \, \sim \, ( \, \ottnt{m_{{\mathrm{1}}}} \, \tau_{{\mathrm{1}}} \, , \, .. \, , \, \ottnt{m_{\ottmv{n}}} \, \tau_{\ottmv{n}} \, )}{}{}{}{\ottcom{MatchList}}\\ 
\ottprodline{|}{ \Gamma  \;  \vdash  \;  \ottnt{d}  : \text{\sf decl} }{}{}{}{\ottcom{Declaration typing}}\\ 
\ottprodline{|}{ \Gamma  \;  \vdash  \;  \ottnt{c}  : \text{\sf comm} }{}{}{}{\ottcom{Command typing}}}

\newcommand{\ottevalXXcomm}{
\ottrulehead{\ottnt{eval\_comm}}{::=}{}\\ 
\ottprodline{|}{\mu \, \ottkw{\{} \, \mathit{x} \, \leftarrow \, \ottnt{v} \, \ottkw{\}} \, \mapsto \, \mu'}{}{}{}{\ottcom{Store Update}}\\ 
\ottprodline{|}{ \langle  \ottnt{c}  ,  \mu  \rangle \mapsto^{ \ottnt{k} } \langle  \ottnt{c'}  ,  \mu'  \rangle }{}{}{}{\ottcom{Many Steps}}\\ 
\ottprodline{|}{ \langle  \ottnt{c}  ;  \mu  \rangle \Rightarrow^{ \ottnt{k} }  \ottnt{tr} }{}{}{}{\ottcom{Trace}}\\ 
\ottprodline{|}{\langle \, \ottnt{c} \, ; \, \mu \, \rangle \, \leadsto \, \mu'}{}{}{}{\ottcom{Full evaluation}}\\ 
\ottprodline{|}{( \, \ottcomp{\mathit{x'_{\ottmv{i}}}:\ottnt{m'_{\ottmv{i}}}\tau'_{\ottmv{i}}}{\ottmv{i}} \, ) \, \ottkw{\#} \, ( \, \ottcomp{\ottnt{e'_{\ottmv{j}}}}{\ottmv{j}} \, ) \, = \, [ \, \ottcomp{\mathit{x_{\ottmv{n}}}:\ottnt{m_{\ottmv{n}}}\tau_{\ottmv{n}}=\ottnt{e_{\ottmv{n}}}}{\ottmv{n}} \, ]}{}{}{}{\ottcom{Compatibility}}\\ 
\ottprodline{|}{ \langle  \ottnt{c}  ,  \mu  \rangle \mapsto \langle  \ottnt{c'}  ,  \mu'  \rangle }{}{}{}{\ottcom{One step evaluation}}\\ 
\ottprodline{|}{\langle \, \ottnt{d} \, , \, \mu \, \rangle \, \mapsto \, \langle \, \ottnt{d'} \, , \, \mu' \, \rangle}{}{}{}{\ottcom{Declaration evaluation}}}

\newcommand{\ottjudgement}{
\ottrulehead{\ottnt{judgement}}{::=}{}\\ 
\ottprodline{|}{\ottnt{eval\_exp}}{}{}{}{}\\ 
\ottprodline{|}{\ottnt{typing}}{}{}{}{}\\ 
\ottprodline{|}{\ottnt{eval\_comm}}{}{}{}{}}

\newcommand{\ottuserXXsyntax}{
\ottrulehead{\ottnt{user\_syntax}}{::=}{}\\ 
\ottprodline{|}{\ottmv{index}}{}{}{}{}\\ 
\ottprodline{|}{\mathit{ident}}{}{}{}{}\\ 
\ottprodline{|}{\ottmv{number}}{}{}{}{}\\ 
\ottprodline{|}{\ottnt{terminals}}{}{}{}{}\\ 
\ottprodline{|}{\ottnt{mode}}{}{}{}{}\\ 
\ottprodline{|}{\ottnt{integer}}{}{}{}{}\\ 
\ottprodline{|}{\ottnt{boolean}}{}{}{}{}\\ 
\ottprodline{|}{\ottnt{exp}}{}{}{}{}\\ 
\ottprodline{|}{\ottnt{store}}{}{}{}{}\\ 
\ottprodline{|}{\ottnt{trace}}{}{}{}{}\\ 
\ottprodline{|}{\ottnt{formula}}{}{}{}{}\\ 
\ottprodline{|}{\ottnt{env}}{}{}{}{}\\ 
\ottprodline{|}{\ottnt{cmd}}{}{}{}{}\\ 
\ottprodline{|}{\ottnt{va}}{}{}{}{}\\ 
\ottprodline{|}{\ottnt{ty}}{}{}{}{}\\ 
\ottprodline{|}{\delta}{}{}{}{}\\ 
\ottprodline{|}{\ottnt{dcl}}{}{}{}{}\\ 
\ottprodline{|}{\ottnt{value}}{}{}{}{}}

\newcommand{\ottgrammar}{\ottgrammartabular{
\ottterminals\ottinterrule
\ottmode\ottinterrule
\ottinteger\ottinterrule
\ottboolean\ottinterrule
\ottexp\ottinterrule
\ottstore\ottinterrule
\otttrace\ottinterrule
\ottformula\ottinterrule
\ottenv\ottinterrule
\ottcmd\ottinterrule
\ottva\ottinterrule
\ottty\ottinterrule
\ottdf\ottinterrule
\ottdcl\ottinterrule
\ottvalue\ottinterrule
\ottevalXXexp\ottinterrule
\otttyping\ottinterrule
\ottevalXXcomm\ottinterrule
\ottjudgement\ottinterrule
\ottuserXXsyntax\ottafterlastrule
}}

% defnss
% defns eval_exp
% defn Fetch
\newcommand{\ottdruleFetchOne}[1]{\ottdrule[#1]{%
}{
( \, \mu \, , \, \mathit{x} \, \leftarrow \, \ottnt{v} \, ) \, ( \, \mathit{x} \, ) \, = \, \ottnt{v}}{%
{\ottdrulename{Fetch1}}{}%
}}

\newcommand{\ottdruleFetchTwo}[1]{\ottdrule[#1]{%
\ottpremise{\mathit{x} \, \neq \, \mathit{x'}}%
\ottpremise{\mu \, ( \, \mathit{x} \, ) \, = \, \ottnt{v}}%
}{
( \, \mu \, , \, \mathit{x'} \, \leftarrow \, \ottnt{v'} \, ) \, ( \, \mathit{x} \, ) \, = \, \ottnt{v}}{%
{\ottdrulename{Fetch2}}{}%
}}

\newcommand{\ottdefnFetch}[1]{\begin{ottdefnblock}[#1]{$\mu \, ( \, \mathit{x} \, ) \, = \, \ottnt{v}$}{\ottcom{Fetch}}
\ottusedrule{\ottdruleFetchOne{}}
\ottusedrule{\ottdruleFetchTwo{}}
\end{ottdefnblock}}

% defn ExpEval
\newcommand{\ottdruleEXXValue}[1]{\ottdrule[#1]{%
}{
 \ottnt{v}  =_ \mu   \ottnt{v} }{%
{\ottdrulename{E\_Value}}{}%
}}

\newcommand{\ottdruleEXXIdent}[1]{\ottdrule[#1]{%
\ottpremise{\mu \, ( \, \mathit{x} \, ) \, = \, \ottnt{v}}%
}{
 \mathit{x}  =_ \mu   \ottnt{v} }{%
{\ottdrulename{E\_Ident}}{}%
}}

\newcommand{\ottdruleEXXPlus}[1]{\ottdrule[#1]{%
\ottpremise{ \ottnt{e_{{\mathrm{1}}}}  =_ \mu   \ottnt{k_{{\mathrm{1}}}} }%
\ottpremise{ \ottnt{e_{{\mathrm{2}}}}  =_ \mu   \ottnt{k_{{\mathrm{2}}}} }%
}{
 \ottnt{e_{{\mathrm{1}}}} \, + \, \ottnt{e_{{\mathrm{2}}}}  =_ \mu   \ottkw{\{} \, \ottnt{k_{{\mathrm{1}}}} \, + \, \ottnt{k_{{\mathrm{2}}}} \, \ottkw{\}} }{%
{\ottdrulename{E\_Plus}}{}%
}}

\newcommand{\ottdruleEXXMinus}[1]{\ottdrule[#1]{%
\ottpremise{ \ottnt{e_{{\mathrm{1}}}}  =_ \mu   \ottnt{k_{{\mathrm{1}}}} }%
\ottpremise{ \ottnt{e_{{\mathrm{2}}}}  =_ \mu   \ottnt{k_{{\mathrm{2}}}} }%
}{
 \ottnt{e_{{\mathrm{1}}}} \, - \, \ottnt{e_{{\mathrm{2}}}}  =_ \mu   \ottkw{\{} \, \ottnt{k_{{\mathrm{1}}}} \, - \, \ottnt{k_{{\mathrm{2}}}} \, \ottkw{\}} }{%
{\ottdrulename{E\_Minus}}{}%
}}

\newcommand{\ottdruleEXXTimes}[1]{\ottdrule[#1]{%
\ottpremise{ \ottnt{e_{{\mathrm{1}}}}  =_ \mu   \ottnt{k_{{\mathrm{1}}}} }%
\ottpremise{ \ottnt{e_{{\mathrm{2}}}}  =_ \mu   \ottnt{k_{{\mathrm{2}}}} }%
}{
 \ottnt{e_{{\mathrm{1}}}} \, \times \, \ottnt{e_{{\mathrm{2}}}}  =_ \mu   \ottkw{\{} \, \ottnt{k_{{\mathrm{1}}}} \, \times \, \ottnt{k_{{\mathrm{2}}}} \, \ottkw{\}} }{%
{\ottdrulename{E\_Times}}{}%
}}

\newcommand{\ottdruleEXXGreater}[1]{\ottdrule[#1]{%
\ottpremise{ \ottnt{e_{{\mathrm{1}}}}  =_ \mu   \ottnt{k_{{\mathrm{1}}}} }%
\ottpremise{ \ottnt{e_{{\mathrm{2}}}}  =_ \mu   \ottnt{k_{{\mathrm{2}}}} }%
}{
  \ottnt{e_{{\mathrm{1}}}}  >  \ottnt{e_{{\mathrm{2}}}}   =_ \mu    \ottnt{k_{{\mathrm{1}}}}  >  \ottnt{k_{{\mathrm{2}}}}  }{%
{\ottdrulename{E\_Greater}}{}%
}}

\newcommand{\ottdruleEXXLess}[1]{\ottdrule[#1]{%
\ottpremise{ \ottnt{e_{{\mathrm{1}}}}  =_ \mu   \ottnt{k_{{\mathrm{1}}}} }%
\ottpremise{ \ottnt{e_{{\mathrm{2}}}}  =_ \mu   \ottnt{k_{{\mathrm{2}}}} }%
}{
  \ottnt{e_{{\mathrm{1}}}}  <  \ottnt{e_{{\mathrm{2}}}}   =_ \mu    \ottnt{k_{{\mathrm{1}}}}  <  \ottnt{k_{{\mathrm{2}}}}  }{%
{\ottdrulename{E\_Less}}{}%
}}

\newcommand{\ottdruleEXXEqual}[1]{\ottdrule[#1]{%
\ottpremise{ \ottnt{e_{{\mathrm{1}}}}  =_ \mu   \ottnt{k_{{\mathrm{1}}}} }%
\ottpremise{ \ottnt{e_{{\mathrm{2}}}}  =_ \mu   \ottnt{k_{{\mathrm{2}}}} }%
}{
 \ottnt{e_{{\mathrm{1}}}} \, = \, \ottnt{e_{{\mathrm{2}}}}  =_ \mu   \ottkw{\{} \, \ottnt{k_{{\mathrm{1}}}} \, = \, \ottnt{k_{{\mathrm{2}}}} \, \ottkw{\}} }{%
{\ottdrulename{E\_Equal}}{}%
}}

\newcommand{\ottdruleEXXAnd}[1]{\ottdrule[#1]{%
\ottpremise{ \ottnt{e_{{\mathrm{1}}}}  =_ \mu   \ottnt{b_{{\mathrm{1}}}} }%
\ottpremise{ \ottnt{e_{{\mathrm{2}}}}  =_ \mu   \ottnt{b_{{\mathrm{2}}}} }%
}{
 \ottnt{e_{{\mathrm{1}}}} \, \ottkw{and} \, \ottnt{e_{{\mathrm{2}}}}  =_ \mu   \ottkw{\{} \, \ottnt{b_{{\mathrm{1}}}} \, \ottkw{and} \, \ottnt{b_{{\mathrm{2}}}} \, \ottkw{\}} }{%
{\ottdrulename{E\_And}}{}%
}}

\newcommand{\ottdruleEXXOr}[1]{\ottdrule[#1]{%
\ottpremise{ \ottnt{e_{{\mathrm{1}}}}  =_ \mu   \ottnt{b_{{\mathrm{1}}}} }%
\ottpremise{ \ottnt{e_{{\mathrm{2}}}}  =_ \mu   \ottnt{b_{{\mathrm{2}}}} }%
}{
 \ottnt{e_{{\mathrm{1}}}} \, \ottkw{or} \, \ottnt{e_{{\mathrm{2}}}}  =_ \mu   \ottkw{\{} \, \ottnt{b_{{\mathrm{1}}}} \, \ottkw{or} \, \ottnt{b_{{\mathrm{2}}}} \, \ottkw{\}} }{%
{\ottdrulename{E\_Or}}{}%
}}

\newcommand{\ottdruleEXXNot}[1]{\ottdrule[#1]{%
\ottpremise{ \ottnt{e}  =_ \mu   \ottnt{b} }%
}{
 \ottkw{not} \, \ottnt{e}  =_ \mu   \ottkw{\{} \, \ottkw{not} \, \ottnt{b} \, \ottkw{\}} }{%
{\ottdrulename{E\_Not}}{}%
}}

\newcommand{\ottdefnExpEval}[1]{\begin{ottdefnblock}[#1]{$ \ottnt{e}  =_ \mu   \ottnt{v} $}{\ottcom{Expression evaluation}}
\ottusedrule{\ottdruleEXXValue{}}
\ottusedrule{\ottdruleEXXIdent{}}
\ottusedrule{\ottdruleEXXPlus{}}
\ottusedrule{\ottdruleEXXMinus{}}
\ottusedrule{\ottdruleEXXTimes{}}
\ottusedrule{\ottdruleEXXGreater{}}
\ottusedrule{\ottdruleEXXLess{}}
\ottusedrule{\ottdruleEXXEqual{}}
\ottusedrule{\ottdruleEXXAnd{}}
\ottusedrule{\ottdruleEXXOr{}}
\ottusedrule{\ottdruleEXXNot{}}
\end{ottdefnblock}}

\newcommand{\ottdefnsevalXXexp}{
\ottdefnFetch{}
\ottdefnExpEval{}}

% defns typing
% defn Lookup
\newcommand{\ottdruleLookupOne}[1]{\ottdrule[#1]{%
}{
 \mathit{ \mathit{x} } \;  \delta  \; \in \;  \Gamma \, , \, \mathit{x} \, \delta }{%
{\ottdrulename{Lookup1}}{}%
}}

\newcommand{\ottdruleLookupTwo}[1]{\ottdrule[#1]{%
\ottpremise{\mathit{x} \, \neq \, \mathit{x'}}%
\ottpremise{ \mathit{ \mathit{x} } \;  \delta  \; \in \;  \Gamma }%
}{
 \mathit{ \mathit{x} } \;  \delta  \; \in \;  \Gamma \, , \, \mathit{x'} \, \delta' }{%
{\ottdrulename{Lookup2}}{}%
}}

\newcommand{\ottdefnLookup}[1]{\begin{ottdefnblock}[#1]{$ \mathit{ \mathit{x} } \;  \delta  \; \in \;  \Gamma $}{\ottcom{Lookup}}
\ottusedrule{\ottdruleLookupOne{}}
\ottusedrule{\ottdruleLookupTwo{}}
\end{ottdefnblock}}

% defn ExpTyping
\newcommand{\ottdruleVar}[1]{\ottdrule[#1]{%
\ottpremise{\ottnt{m} \, \neq \, \ottkw{out}}%
\ottpremise{ \mathit{ \mathit{x} } \;  : \, \ottnt{m} \, \tau  \; \in \;  \Gamma }%
}{
\Gamma \, \vdash \, \mathit{x} \, : \, \tau}{%
{\ottdrulename{Var}}{}%
}}

\newcommand{\ottdruleIntCst}[1]{\ottdrule[#1]{%
}{
\Gamma \, \vdash \, \ottmv{q} \, : \, \ottkw{int}}{%
{\ottdrulename{IntCst}}{}%
}}

\newcommand{\ottdruleBoolTrue}[1]{\ottdrule[#1]{%
}{
\Gamma \, \vdash \, \ottkw{true} \, : \, \ottkw{bool}}{%
{\ottdrulename{BoolTrue}}{}%
}}

\newcommand{\ottdruleBoolFalse}[1]{\ottdrule[#1]{%
}{
\Gamma \, \vdash \, \ottkw{false} \, : \, \ottkw{bool}}{%
{\ottdrulename{BoolFalse}}{}%
}}

\newcommand{\ottdrulePlus}[1]{\ottdrule[#1]{%
\ottpremise{\Gamma \, \vdash \, \ottnt{e_{{\mathrm{1}}}} \, : \, \ottkw{int}}%
\ottpremise{\Gamma \, \vdash \, \ottnt{e_{{\mathrm{2}}}} \, : \, \ottkw{int}}%
}{
\Gamma \, \vdash \, \ottnt{e_{{\mathrm{1}}}} \, + \, \ottnt{e_{{\mathrm{2}}}} \, : \, \ottkw{int}}{%
{\ottdrulename{Plus}}{}%
}}

\newcommand{\ottdruleMinus}[1]{\ottdrule[#1]{%
\ottpremise{\Gamma \, \vdash \, \ottnt{e_{{\mathrm{1}}}} \, : \, \ottkw{int}}%
\ottpremise{\Gamma \, \vdash \, \ottnt{e_{{\mathrm{2}}}} \, : \, \ottkw{int}}%
}{
\Gamma \, \vdash \, \ottnt{e_{{\mathrm{1}}}} \, - \, \ottnt{e_{{\mathrm{2}}}} \, : \, \ottkw{int}}{%
{\ottdrulename{Minus}}{}%
}}

\newcommand{\ottdruleTimes}[1]{\ottdrule[#1]{%
\ottpremise{\Gamma \, \vdash \, \ottnt{e_{{\mathrm{1}}}} \, : \, \ottkw{int}}%
\ottpremise{\Gamma \, \vdash \, \ottnt{e_{{\mathrm{2}}}} \, : \, \ottkw{int}}%
}{
\Gamma \, \vdash \, \ottnt{e_{{\mathrm{1}}}} \, \times \, \ottnt{e_{{\mathrm{2}}}} \, : \, \ottkw{int}}{%
{\ottdrulename{Times}}{}%
}}

\newcommand{\ottdruleEqual}[1]{\ottdrule[#1]{%
\ottpremise{\Gamma \, \vdash \, \ottnt{e_{{\mathrm{1}}}} \, : \, \tau}%
\ottpremise{\Gamma \, \vdash \, \ottnt{e_{{\mathrm{2}}}} \, : \, \tau}%
}{
\Gamma \, \vdash \, \ottnt{e_{{\mathrm{1}}}} \, = \, \ottnt{e_{{\mathrm{2}}}} \, : \, \ottkw{bool}}{%
{\ottdrulename{Equal}}{}%
}}

\newcommand{\ottdruleGreater}[1]{\ottdrule[#1]{%
\ottpremise{\Gamma \, \vdash \, \ottnt{e_{{\mathrm{1}}}} \, : \, \ottkw{int}}%
\ottpremise{\Gamma \, \vdash \, \ottnt{e_{{\mathrm{2}}}} \, : \, \ottkw{int}}%
}{
\Gamma \, \vdash \,  \ottnt{e_{{\mathrm{1}}}}  >  \ottnt{e_{{\mathrm{2}}}}  \, : \, \ottkw{bool}}{%
{\ottdrulename{Greater}}{}%
}}

\newcommand{\ottdruleLess}[1]{\ottdrule[#1]{%
\ottpremise{\Gamma \, \vdash \, \ottnt{e_{{\mathrm{1}}}} \, : \, \ottkw{int}}%
\ottpremise{\Gamma \, \vdash \, \ottnt{e_{{\mathrm{2}}}} \, : \, \ottkw{int}}%
}{
\Gamma \, \vdash \,  \ottnt{e_{{\mathrm{1}}}}  <  \ottnt{e_{{\mathrm{2}}}}  \, : \, \ottkw{bool}}{%
{\ottdrulename{Less}}{}%
}}

\newcommand{\ottdruleAnd}[1]{\ottdrule[#1]{%
\ottpremise{\Gamma \, \vdash \, \ottnt{e_{{\mathrm{1}}}} \, : \, \ottkw{bool}}%
\ottpremise{\Gamma \, \vdash \, \ottnt{e_{{\mathrm{2}}}} \, : \, \ottkw{bool}}%
}{
\Gamma \, \vdash \, \ottnt{e_{{\mathrm{1}}}} \, \ottkw{and} \, \ottnt{e_{{\mathrm{2}}}} \, : \, \ottkw{bool}}{%
{\ottdrulename{And}}{}%
}}

\newcommand{\ottdruleOr}[1]{\ottdrule[#1]{%
\ottpremise{\Gamma \, \vdash \, \ottnt{e_{{\mathrm{1}}}} \, : \, \ottkw{bool}}%
\ottpremise{\Gamma \, \vdash \, \ottnt{e_{{\mathrm{2}}}} \, : \, \ottkw{bool}}%
}{
\Gamma \, \vdash \, \ottnt{e_{{\mathrm{1}}}} \, \ottkw{or} \, \ottnt{e_{{\mathrm{2}}}} \, : \, \ottkw{bool}}{%
{\ottdrulename{Or}}{}%
}}

\newcommand{\ottdruleNot}[1]{\ottdrule[#1]{%
\ottpremise{\Gamma \, \vdash \, \ottnt{e} \, : \, \ottkw{bool}}%
}{
\Gamma \, \vdash \, \ottkw{not} \, \ottnt{e} \, : \, \ottkw{bool}}{%
{\ottdrulename{Not}}{}%
}}

\newcommand{\ottdefnExpTyping}[1]{\begin{ottdefnblock}[#1]{$\Gamma \, \vdash \, \ottnt{e} \, : \, \tau$}{\ottcom{Expression typing}}
\ottusedrule{\ottdruleVar{}}
\ottusedrule{\ottdruleIntCst{}}
\ottusedrule{\ottdruleBoolTrue{}}
\ottusedrule{\ottdruleBoolFalse{}}
\ottusedrule{\ottdrulePlus{}}
\ottusedrule{\ottdruleMinus{}}
\ottusedrule{\ottdruleTimes{}}
\ottusedrule{\ottdruleEqual{}}
\ottusedrule{\ottdruleGreater{}}
\ottusedrule{\ottdruleLess{}}
\ottusedrule{\ottdruleAnd{}}
\ottusedrule{\ottdruleOr{}}
\ottusedrule{\ottdruleNot{}}
\end{ottdefnblock}}

% defn LookupD
\newcommand{\ottdruleLookupDOne}[1]{\ottdrule[#1]{%
}{
 \delta  \; \in \;  \Gamma \, , \, \mathit{x} \, \delta }{%
{\ottdrulename{LookupD1}}{}%
}}

\newcommand{\ottdruleLookupDTwo}[1]{\ottdrule[#1]{%
\ottpremise{\delta \, \neq \, \delta'}%
\ottpremise{ \delta  \; \in \;  \Gamma }%
}{
 \delta  \; \in \;  \Gamma \, , \, \mathit{x} \, \delta' }{%
{\ottdrulename{LookupD2}}{}%
}}

\newcommand{\ottdefnLookupD}[1]{\begin{ottdefnblock}[#1]{$ \delta  \; \in \;  \Gamma $}{\ottcom{LookupD}}
\ottusedrule{\ottdruleLookupDOne{}}
\ottusedrule{\ottdruleLookupDTwo{}}
\end{ottdefnblock}}

% defn Match
\newcommand{\ottdruleMatchOne}[1]{\ottdrule[#1]{%
\ottpremise{\Gamma \, \vdash \, \ottnt{e} \, : \, \tau}%
}{
\Gamma \, \vdash \, \ottnt{e} \, \sim \, \ottkw{in} \, \tau}{%
{\ottdrulename{Match1}}{}%
}}

\newcommand{\ottdruleMatchTwo}[1]{\ottdrule[#1]{%
\ottpremise{ \mathit{ \mathit{x} } \;  : \, \ottnt{m} \, \tau  \; \in \;  \Gamma }%
\ottpremise{\ottnt{m} \, \neq \, \ottkw{in}}%
}{
\Gamma \, \vdash \, \mathit{x} \, \sim \, \ottkw{out} \, \tau}{%
{\ottdrulename{Match2}}{}%
}}

\newcommand{\ottdruleMatchThree}[1]{\ottdrule[#1]{%
\ottpremise{ \mathit{ \mathit{x} } \;  : \, \ottkw{in} \, \ottkw{out} \, \tau  \; \in \;  \Gamma }%
}{
\Gamma \, \vdash \, \mathit{x} \, \sim \, \ottkw{in} \, \ottkw{out} \, \tau}{%
{\ottdrulename{Match3}}{}%
}}

\newcommand{\ottdefnMatch}[1]{\begin{ottdefnblock}[#1]{$\Gamma \, \vdash \, \ottnt{e} \, \sim \, \ottnt{m} \, \tau$}{\ottcom{Match}}
\ottusedrule{\ottdruleMatchOne{}}
\ottusedrule{\ottdruleMatchTwo{}}
\ottusedrule{\ottdruleMatchThree{}}
\end{ottdefnblock}}

% defn MatchList
\newcommand{\ottdruleMatchListOne}[1]{\ottdrule[#1]{%
}{
\Gamma \, \vdash \, ( \, \, \, ) \, \sim \, ( \, \, \, )}{%
{\ottdrulename{MatchList1}}{}%
}}

\newcommand{\ottdruleMatchListTwo}[1]{\ottdrule[#1]{%
\ottpremise{\Gamma \, \vdash \, \ottnt{e} \, \sim \, \ottnt{m} \, \tau}%
\ottpremise{\Gamma \, \vdash \, ( \, \ottnt{e_{{\mathrm{1}}}} \, , \, .. \, , \, \ottnt{e_{\ottmv{l}}} \, ) \, \sim \, ( \, \ottnt{m_{{\mathrm{1}}}} \, \tau_{{\mathrm{1}}} \, , \, .. \, , \, \ottnt{m_{\ottmv{n}}} \, \tau_{\ottmv{n}} \, )}%
}{
\Gamma \, \vdash \, ( \, \ottnt{e} \, , \, \ottnt{e_{{\mathrm{1}}}} \, , \, .. \, , \, \ottnt{e_{\ottmv{l}}} \, ) \, \sim \, ( \, \ottnt{m} \, \tau \, , \, \ottnt{m_{{\mathrm{1}}}} \, \tau_{{\mathrm{1}}} \, , \, .. \, , \, \ottnt{m_{\ottmv{n}}} \, \tau_{\ottmv{n}} \, )}{%
{\ottdrulename{MatchList2}}{}%
}}

\newcommand{\ottdefnMatchList}[1]{\begin{ottdefnblock}[#1]{$\Gamma \, \vdash \, ( \, \ottnt{e_{{\mathrm{1}}}} \, , \, .. \, , \, \ottnt{e_{\ottmv{l}}} \, ) \, \sim \, ( \, \ottnt{m_{{\mathrm{1}}}} \, \tau_{{\mathrm{1}}} \, , \, .. \, , \, \ottnt{m_{\ottmv{n}}} \, \tau_{\ottmv{n}} \, )$}{\ottcom{MatchList}}
\ottusedrule{\ottdruleMatchListOne{}}
\ottusedrule{\ottdruleMatchListTwo{}}
\end{ottdefnblock}}

% defn DeclTyping
\newcommand{\ottdruleEmpty}[1]{\ottdrule[#1]{%
}{
 \Gamma  \;  \vdash  \;  \ottkw{begin} \, \ottkw{end}  : \text{\sf decl} }{%
{\ottdrulename{Empty}}{}%
}}

\newcommand{\ottdruleBlock}[1]{\ottdrule[#1]{%
\ottpremise{ \Gamma  \;  \vdash  \;  \ottnt{c}  : \text{\sf comm} }%
}{
 \Gamma  \;  \vdash  \;  \ottkw{begin} \, \ottnt{c} \, ; \, \ottkw{end}  : \text{\sf decl} }{%
{\ottdrulename{Block}}{}%
}}

\newcommand{\ottdruleUninitVar}[1]{\ottdrule[#1]{%
\ottpremise{ \Gamma \, , \, \mathit{x} \, : \, \ottkw{in} \, \ottkw{out} \, \tau  \;  \vdash  \;  \ottnt{d}  : \text{\sf decl} }%
}{
 \Gamma  \;  \vdash  \;  \mathit{x} \, : \, \tau \, ; \, \ottnt{d}  : \text{\sf decl} }{%
{\ottdrulename{UninitVar}}{}%
}}

\newcommand{\ottdruleInitVar}[1]{\ottdrule[#1]{%
\ottpremise{\Gamma \, \vdash \, \ottnt{e} \, : \, \tau}%
\ottpremise{ \Gamma \, , \, \mathit{x} \, : \, \ottkw{in} \, \ottkw{out} \, \tau  \;  \vdash  \;  \ottnt{d}  : \text{\sf decl} }%
}{
 \Gamma  \;  \vdash  \;  \mathit{x} \, : \, \tau \, := \, \ottnt{e} \, ; \, \ottnt{d}  : \text{\sf decl} }{%
{\ottdrulename{InitVar}}{}%
}}

\newcommand{\ottdruleConstant}[1]{\ottdrule[#1]{%
\ottpremise{\Gamma \, \vdash \, \ottnt{e} \, : \, \tau}%
\ottpremise{ \Gamma \, , \, \mathit{x} \, : \, \ottkw{in} \, \tau  \;  \vdash  \;  \ottnt{d}  : \text{\sf decl} }%
}{
 \Gamma  \;  \vdash  \;  \mathit{x} \, : \, \ottkw{constant} \, \tau \, := \, \ottnt{e} \, ; \, \ottnt{d}  : \text{\sf decl} }{%
{\ottdrulename{Constant}}{}%
}}

\newcommand{\ottdruleProc}[1]{\ottdrule[#1]{%
\ottpremise{  \Gamma \, , \, \mathit{x_{{\mathrm{1}}}} \, : \, \ottnt{m_{{\mathrm{1}}}} \, \tau_{{\mathrm{1}}} \, , \, .. \, , \, \mathit{x_{\ottmv{n}}} \, : \, \ottnt{m_{\ottmv{n}}} \, \tau_{\ottmv{n}}   \;  \vdash  \;  \ottnt{d_{{\mathrm{1}}}}  : \text{\sf decl} }%
\ottpremise{ \Gamma \, , \, \mathit{p} \, : \, \ottkw{in} \, \ottkw{proc} \, ( \, \ottnt{m_{{\mathrm{1}}}} \, \tau_{{\mathrm{1}}} \, , \, .. \, , \, \ottnt{m_{\ottmv{n}}} \, \tau_{\ottmv{n}} \, )  \;  \vdash  \;  \ottnt{d_{{\mathrm{2}}}}  : \text{\sf decl} }%
}{
 \Gamma  \;  \vdash  \;  \ottkw{procedure} \, \mathit{p} \, ( \, \mathit{x_{{\mathrm{1}}}} \, : \, \ottnt{m_{{\mathrm{1}}}} \, \tau_{{\mathrm{1}}} \, ; \, .. \, ; \, \mathit{x_{\ottmv{n}}} \, : \, \ottnt{m_{\ottmv{n}}} \, \tau_{\ottmv{n}} \, ) \, \ottkw{is} \, \ottnt{d_{{\mathrm{1}}}} \, ; \, \ottnt{d_{{\mathrm{2}}}}  : \text{\sf decl} }{%
{\ottdrulename{Proc}}{}%
}}

\newcommand{\ottdefnDeclTyping}[1]{\begin{ottdefnblock}[#1]{$ \Gamma  \;  \vdash  \;  \ottnt{d}  : \text{\sf decl} $}{\ottcom{Declaration typing}}
\ottusedrule{\ottdruleEmpty{}}
\ottusedrule{\ottdruleBlock{}}
\ottusedrule{\ottdruleUninitVar{}}
\ottusedrule{\ottdruleInitVar{}}
\ottusedrule{\ottdruleConstant{}}
\ottusedrule{\ottdruleProc{}}
\end{ottdefnblock}}

% defn CommTyping
\newcommand{\ottdruleNull}[1]{\ottdrule[#1]{%
}{
 \Gamma  \;  \vdash  \;  \ottkw{null}  : \text{\sf comm} }{%
{\ottdrulename{Null}}{}%
}}

\newcommand{\ottdruleSeq}[1]{\ottdrule[#1]{%
\ottpremise{ \Gamma  \;  \vdash  \;  \ottnt{c_{{\mathrm{1}}}}  : \text{\sf comm} }%
\ottpremise{ \Gamma  \;  \vdash  \;  \ottnt{c_{{\mathrm{2}}}}  : \text{\sf comm} }%
}{
 \Gamma  \;  \vdash  \;  \ottnt{c_{{\mathrm{1}}}} \, ; \, \ottnt{c_{{\mathrm{2}}}}  : \text{\sf comm} }{%
{\ottdrulename{Seq}}{}%
}}

\newcommand{\ottdruleAssign}[1]{\ottdrule[#1]{%
\ottpremise{\ottnt{m} \, \neq \, \ottkw{in}}%
\ottpremise{ \mathit{ \mathit{x} } \;  : \, \ottnt{m} \, \tau  \; \in \;  \Gamma }%
\ottpremise{\Gamma \, \vdash \, \ottnt{e} \, : \, \tau}%
}{
 \Gamma  \;  \vdash  \;  \mathit{x} \, := \, \ottnt{e}  : \text{\sf comm} }{%
{\ottdrulename{Assign}}{}%
}}

\newcommand{\ottdruleIfThenElse}[1]{\ottdrule[#1]{%
\ottpremise{\Gamma \, \vdash \, \ottnt{e} \, : \, \ottkw{bool}}%
\ottpremise{ \Gamma  \;  \vdash  \;  \ottnt{c_{{\mathrm{1}}}}  : \text{\sf comm} }%
\ottpremise{ \Gamma  \;  \vdash  \;  \ottnt{c_{{\mathrm{2}}}}  : \text{\sf comm} }%
}{
 \Gamma  \;  \vdash  \;  \ottkw{if} \, \ottnt{e} \, \ottkw{then} \, \ottnt{c_{{\mathrm{1}}}} \, ; \, \ottkw{else} \, \ottnt{c_{{\mathrm{2}}}} \, ; \, \ottkw{end} \, \ottkw{if}  : \text{\sf comm} }{%
{\ottdrulename{IfThenElse}}{}%
}}

\newcommand{\ottdruleWhile}[1]{\ottdrule[#1]{%
\ottpremise{\Gamma \, \vdash \, \ottnt{e} \, : \, \ottkw{bool}}%
\ottpremise{ \Gamma  \;  \vdash  \;  \ottnt{c}  : \text{\sf comm} }%
}{
 \Gamma  \;  \vdash  \;  \ottkw{while} \, \ottnt{e} \, \ottkw{loop} \, \ottnt{c} \, ; \, \ottkw{end} \, \ottkw{loop}  : \text{\sf comm} }{%
{\ottdrulename{While}}{}%
}}

\newcommand{\ottdruleFor}[1]{\ottdrule[#1]{%
\ottpremise{\Gamma \, \vdash \, \ottnt{e} \, : \, \ottkw{int}}%
\ottpremise{\Gamma \, \vdash \, \ottnt{e'} \, : \, \ottkw{int}}%
\ottpremise{ \Gamma \, , \, \mathit{x} \, : \, \ottkw{in} \, \ottkw{int}  \;  \vdash  \;  \ottnt{c}  : \text{\sf comm} }%
}{
 \Gamma  \;  \vdash  \;  \ottkw{for} \, \mathit{x} \, \ottkw{in} \, \ottnt{e} \, . \, . \, \ottnt{e'} \, \ottkw{loop} \, \ottnt{c} \, ; \, \ottkw{end} \, \ottkw{loop}  : \text{\sf comm} }{%
{\ottdrulename{For}}{}%
}}

\newcommand{\ottdruleDecl}[1]{\ottdrule[#1]{%
\ottpremise{ \Gamma  \;  \vdash  \;  \ottnt{d}  : \text{\sf decl} }%
}{
 \Gamma  \;  \vdash  \;  \ottkw{declare} \, \ottnt{d}  : \text{\sf comm} }{%
{\ottdrulename{Decl}}{}%
}}

\newcommand{\ottdruleProcCall}[1]{\ottdrule[#1]{%
\ottpremise{\Gamma \, \vdash \, \ottnt{e} \, : \, \ottkw{proc} \, ( \, \ottnt{m_{{\mathrm{1}}}} \, \tau_{{\mathrm{1}}} \, , \, .. \, , \, \ottnt{m_{\ottmv{n}}} \, \tau_{\ottmv{n}} \, )}%
\ottpremise{\Gamma \, \vdash \, ( \, \ottnt{e_{{\mathrm{1}}}} \, , \, .. \, , \, \ottnt{e_{\ottmv{l}}} \, ) \, \sim \, ( \, \ottnt{m_{{\mathrm{1}}}} \, \tau_{{\mathrm{1}}} \, , \, .. \, , \, \ottnt{m_{\ottmv{n}}} \, \tau_{\ottmv{n}} \, )}%
}{
 \Gamma  \;  \vdash  \;  \ottnt{e} \, ( \, \ottnt{e_{{\mathrm{1}}}} \, , \, .. \, , \, \ottnt{e_{\ottmv{l}}} \, )  : \text{\sf comm} }{%
{\ottdrulename{ProcCall}}{}%
}}

\newcommand{\ottdefnCommTyping}[1]{\begin{ottdefnblock}[#1]{$ \Gamma  \;  \vdash  \;  \ottnt{c}  : \text{\sf comm} $}{\ottcom{Command typing}}
\ottusedrule{\ottdruleNull{}}
\ottusedrule{\ottdruleSeq{}}
\ottusedrule{\ottdruleAssign{}}
\ottusedrule{\ottdruleIfThenElse{}}
\ottusedrule{\ottdruleWhile{}}
\ottusedrule{\ottdruleFor{}}
\ottusedrule{\ottdruleDecl{}}
\ottusedrule{\ottdruleProcCall{}}
\end{ottdefnblock}}

\newcommand{\ottdefnstyping}{
\ottdefnLookup{}
\ottdefnExpTyping{}
\ottdefnLookupD{}
\ottdefnMatch{}
\ottdefnMatchList{}
\ottdefnDeclTyping{}
\ottdefnCommTyping{}}

% defns eval_comm
% defn StoreUpdate
\newcommand{\ottdruleUpdateOne}[1]{\ottdrule[#1]{%
}{
( \, \mu \, , \, \mathit{x} \, \leftarrow \, \ottnt{v} \, ) \, \ottkw{\{} \, \mathit{x} \, \leftarrow \, \ottnt{v'} \, \ottkw{\}} \, \mapsto \, ( \, \mu \, , \, \mathit{x} \, \leftarrow \, \ottnt{v'} \, )}{%
{\ottdrulename{Update1}}{}%
}}

\newcommand{\ottdruleUpdateTwo}[1]{\ottdrule[#1]{%
\ottpremise{\mathit{x} \, \neq \, \mathit{x'}}%
\ottpremise{\mu \, \ottkw{\{} \, \mathit{x} \, \leftarrow \, \ottnt{v'} \, \ottkw{\}} \, \mapsto \, \mu'}%
}{
( \, \mu \, , \, \mathit{x'} \, \leftarrow \, \ottnt{v} \, ) \, \ottkw{\{} \, \mathit{x} \, \leftarrow \, \ottnt{v'} \, \ottkw{\}} \, \mapsto \, ( \, \mu' \, , \, \mathit{x'} \, \leftarrow \, \ottnt{v} \, )}{%
{\ottdrulename{Update2}}{}%
}}

\newcommand{\ottdefnStoreUpdate}[1]{\begin{ottdefnblock}[#1]{$\mu \, \ottkw{\{} \, \mathit{x} \, \leftarrow \, \ottnt{v} \, \ottkw{\}} \, \mapsto \, \mu'$}{\ottcom{Store Update}}
\ottusedrule{\ottdruleUpdateOne{}}
\ottusedrule{\ottdruleUpdateTwo{}}
\end{ottdefnblock}}

% defn ManySteps
\newcommand{\ottdruleManyStepsOne}[1]{\ottdrule[#1]{%
}{
 \langle  \ottnt{c}  ,  \mu  \rangle \mapsto^{ 0 } \langle  \ottnt{c}  ,  \mu  \rangle }{%
{\ottdrulename{ManySteps1}}{}%
}}

\newcommand{\ottdruleManyStepsTwo}[1]{\ottdrule[#1]{%
}{
 \langle  \ottkw{null}  ,  \mu  \rangle \mapsto^{ \ottnt{k} } \langle  \ottkw{null}  ,  \mu  \rangle }{%
{\ottdrulename{ManySteps2}}{}%
}}

\newcommand{\ottdruleManyStepsThree}[1]{\ottdrule[#1]{%
\ottpremise{ \langle  \ottnt{c}  ,  \mu  \rangle \mapsto \langle  \ottnt{c'}  ,  \mu'  \rangle }%
\ottpremise{ \langle  \ottnt{c'}  ,  \mu'  \rangle \mapsto^{ \ottkw{\{} \, \ottnt{k} \, - \, 1 \, \ottkw{\}} } \langle  \ottnt{c''}  ,  \mu''  \rangle }%
}{
 \langle  \ottnt{c}  ,  \mu  \rangle \mapsto^{ \ottnt{k} } \langle  \ottnt{c''}  ,  \mu''  \rangle }{%
{\ottdrulename{ManySteps3}}{}%
}}

\newcommand{\ottdefnManySteps}[1]{\begin{ottdefnblock}[#1]{$ \langle  \ottnt{c}  ,  \mu  \rangle \mapsto^{ \ottnt{k} } \langle  \ottnt{c'}  ,  \mu'  \rangle $}{\ottcom{Many Steps}}
\ottusedrule{\ottdruleManyStepsOne{}}
\ottusedrule{\ottdruleManyStepsTwo{}}
\ottusedrule{\ottdruleManyStepsThree{}}
\end{ottdefnblock}}

% defn Trace
\newcommand{\ottdruleTraceOne}[1]{\ottdrule[#1]{%
}{
 \langle  \ottnt{c}  ;  \mu  \rangle \Rightarrow^{ 0 }  [ \, \, \, ] }{%
{\ottdrulename{Trace1}}{}%
}}

\newcommand{\ottdruleTraceTwo}[1]{\ottdrule[#1]{%
}{
 \langle  \ottkw{null}  ;  \mu  \rangle \Rightarrow^{ \ottnt{k} }  [ \, \, \, ] }{%
{\ottdrulename{Trace2}}{}%
}}

\newcommand{\ottdruleTraceThree}[1]{\ottdrule[#1]{%
\ottpremise{ \langle  \ottnt{c}  ,  \mu  \rangle \mapsto \langle  \ottnt{c'}  ,  \mu'  \rangle }%
\ottpremise{ \langle  \ottnt{c'}  ;  \mu'  \rangle \Rightarrow^{ \ottkw{\{} \, \ottnt{k} \, - \, 1 \, \ottkw{\}} }  [ \, \langle \, \ottnt{c'_{{\mathrm{1}}}} \, , \, \mu'_{{\mathrm{1}}} \, \rangle \, .. \, \langle \, \ottnt{c'_{\ottmv{n}}} \, , \, \mu'_{\ottmv{n}} \, \rangle \, ] }%
}{
 \langle  \ottnt{c}  ;  \mu  \rangle \Rightarrow^{ \ottnt{k} }  [ \, \langle \, \ottnt{c'} \, , \, \mu' \, \rangle \, \langle \, \ottnt{c'_{{\mathrm{1}}}} \, , \, \mu'_{{\mathrm{1}}} \, \rangle \, .. \, \langle \, \ottnt{c'_{\ottmv{n}}} \, , \, \mu'_{\ottmv{n}} \, \rangle \, ] }{%
{\ottdrulename{Trace3}}{}%
}}

\newcommand{\ottdefnTrace}[1]{\begin{ottdefnblock}[#1]{$ \langle  \ottnt{c}  ;  \mu  \rangle \Rightarrow^{ \ottnt{k} }  \ottnt{tr} $}{\ottcom{Trace}}
\ottusedrule{\ottdruleTraceOne{}}
\ottusedrule{\ottdruleTraceTwo{}}
\ottusedrule{\ottdruleTraceThree{}}
\end{ottdefnblock}}

% defn FullEvaluation
\newcommand{\ottdruleEvalOne}[1]{\ottdrule[#1]{%
}{
\langle \, \ottkw{null} \, ; \, \mu \, \rangle \, \leadsto \, \mu}{%
{\ottdrulename{Eval1}}{}%
}}

\newcommand{\ottdruleEvalTwo}[1]{\ottdrule[#1]{%
\ottpremise{ \langle  \ottnt{c}  ,  \mu  \rangle \mapsto \langle  \ottnt{c'}  ,  \mu'  \rangle }%
\ottpremise{\langle \, \ottnt{c'} \, ; \, \mu' \, \rangle \, \leadsto \, \mu''}%
}{
\langle \, \ottnt{c} \, ; \, \mu \, \rangle \, \leadsto \, \mu''}{%
{\ottdrulename{Eval2}}{}%
}}

\newcommand{\ottdefnFullEvaluation}[1]{\begin{ottdefnblock}[#1]{$\langle \, \ottnt{c} \, ; \, \mu \, \rangle \, \leadsto \, \mu'$}{\ottcom{Full evaluation}}
\ottusedrule{\ottdruleEvalOne{}}
\ottusedrule{\ottdruleEvalTwo{}}
\end{ottdefnblock}}

% defn Compat
\newcommand{\ottdruleEXXCompatOne}[1]{\ottdrule[#1]{%
}{
( \, \, \, ) \, \ottkw{\#} \, ( \, \, \, ) \, = \, [ \, \, \, ]}{%
{\ottdrulename{E\_Compat1}}{}%
}}

\newcommand{\ottdruleEXXCompatTwo}[1]{\ottdrule[#1]{%
\ottpremise{( \, \ottcomp{\mathit{x'_{\ottmv{i}}}:\ottnt{m'_{\ottmv{i}}}\tau'_{\ottmv{i}}}{\ottmv{i}} \, ) \, \ottkw{\#} \, ( \, \ottcomp{\ottnt{e'_{\ottmv{j}}}}{\ottmv{j}} \, ) \, = \, [ \, \ottcomp{\mathit{x_{\ottmv{n}}}:\ottnt{m_{\ottmv{n}}}\tau_{\ottmv{n}}=\ottnt{e_{\ottmv{n}}}}{\ottmv{n}} \, ]}%
}{
( \, \mathit{x} \, : \, \ottnt{m} \, \tau \, \ottcomp{\mathit{x'_{\ottmv{i}}}:\ottnt{m'_{\ottmv{i}}}\tau'_{\ottmv{i}}}{\ottmv{i}} \, ) \, \ottkw{\#} \, ( \, \ottnt{e} \, \ottcomp{\ottnt{e'_{\ottmv{j}}}}{\ottmv{j}} \, ) \, = \, [ \, \mathit{x} \, : \, \ottnt{m} \, \tau \, = \, \ottnt{e} \, \ottcomp{\mathit{x_{\ottmv{n}}}:\ottnt{m_{\ottmv{n}}}\tau_{\ottmv{n}}=\ottnt{e_{\ottmv{n}}}}{\ottmv{n}} \, ]}{%
{\ottdrulename{E\_Compat2}}{}%
}}

\newcommand{\ottdefnCompat}[1]{\begin{ottdefnblock}[#1]{$( \, \ottcomp{\mathit{x'_{\ottmv{i}}}:\ottnt{m'_{\ottmv{i}}}\tau'_{\ottmv{i}}}{\ottmv{i}} \, ) \, \ottkw{\#} \, ( \, \ottcomp{\ottnt{e'_{\ottmv{j}}}}{\ottmv{j}} \, ) \, = \, [ \, \ottcomp{\mathit{x_{\ottmv{n}}}:\ottnt{m_{\ottmv{n}}}\tau_{\ottmv{n}}=\ottnt{e_{\ottmv{n}}}}{\ottmv{n}} \, ]$}{\ottcom{Compatibility}}
\ottusedrule{\ottdruleEXXCompatOne{}}
\ottusedrule{\ottdruleEXXCompatTwo{}}
\end{ottdefnblock}}

% defn OneStep
\newcommand{\ottdruleEXXNull}[1]{\ottdrule[#1]{%
}{
 \langle  ( \, \ottkw{null} \, ; \, \ottnt{c} \, )  ,  \mu  \rangle \mapsto \langle  \ottnt{c}  ,  \mu  \rangle }{%
{\ottdrulename{E\_Null}}{}%
}}

\newcommand{\ottdruleEXXSeq}[1]{\ottdrule[#1]{%
\ottpremise{ \langle  \ottnt{c_{{\mathrm{1}}}}  ,  \mu  \rangle \mapsto \langle  \ottnt{c'_{{\mathrm{1}}}}  ,  \mu'  \rangle }%
}{
 \langle  ( \, \ottnt{c_{{\mathrm{1}}}} \, ; \, \ottnt{c_{{\mathrm{2}}}} \, )  ,  \mu  \rangle \mapsto \langle  ( \, \ottnt{c'_{{\mathrm{1}}}} \, ; \, \ottnt{c_{{\mathrm{2}}}} \, )  ,  \mu'  \rangle }{%
{\ottdrulename{E\_Seq}}{}%
}}

\newcommand{\ottdruleEXXAssign}[1]{\ottdrule[#1]{%
\ottpremise{ \ottnt{e}  =_ \mu   \ottnt{v} }%
\ottpremise{\mu \, \ottkw{\{} \, \mathit{x} \, \leftarrow \, \ottnt{v} \, \ottkw{\}} \, \mapsto \, \mu'}%
}{
 \langle  ( \, \mathit{x} \, := \, \ottnt{e} \, )  ,  \mu  \rangle \mapsto \langle  \ottkw{null}  ,  \mu'  \rangle }{%
{\ottdrulename{E\_Assign}}{}%
}}

\newcommand{\ottdruleEXXIfThenElseOne}[1]{\ottdrule[#1]{%
\ottpremise{ \ottnt{e}  =_ \mu   \ottkw{true} }%
}{
 \langle  ( \, \ottkw{if} \, \ottnt{e} \, \ottkw{then} \, \ottnt{c_{{\mathrm{1}}}} \, ; \, \ottkw{else} \, \ottnt{c_{{\mathrm{2}}}} \, ; \, \ottkw{end} \, \ottkw{if} \, )  ,  \mu  \rangle \mapsto \langle  \ottnt{c_{{\mathrm{1}}}}  ,  \mu  \rangle }{%
{\ottdrulename{E\_IfThenElse1}}{}%
}}

\newcommand{\ottdruleEXXIfThenElseTwo}[1]{\ottdrule[#1]{%
\ottpremise{ \ottnt{e}  =_ \mu   \ottkw{false} }%
}{
 \langle  ( \, \ottkw{if} \, \ottnt{e} \, \ottkw{then} \, \ottnt{c_{{\mathrm{1}}}} \, ; \, \ottkw{else} \, \ottnt{c_{{\mathrm{2}}}} \, ; \, \ottkw{end} \, \ottkw{if} \, )  ,  \mu  \rangle \mapsto \langle  \ottnt{c_{{\mathrm{2}}}}  ,  \mu  \rangle }{%
{\ottdrulename{E\_IfThenElse2}}{}%
}}

\newcommand{\ottdruleEXXWhileOne}[1]{\ottdrule[#1]{%
\ottpremise{ \ottnt{e}  =_ \mu   \ottkw{false} }%
}{
 \langle  ( \, \ottkw{while} \, \ottnt{e} \, \ottkw{loop} \, \ottnt{c} \, ; \, \ottkw{end} \, \ottkw{loop} \, )  ,  \mu  \rangle \mapsto \langle  \ottkw{null}  ,  \mu  \rangle }{%
{\ottdrulename{E\_While1}}{}%
}}

\newcommand{\ottdruleEXXWhileTwo}[1]{\ottdrule[#1]{%
\ottpremise{ \ottnt{e}  =_ \mu   \ottkw{true} }%
}{
 \langle  ( \, \ottkw{while} \, \ottnt{e} \, \ottkw{loop} \, \ottnt{c} \, ; \, \ottkw{end} \, \ottkw{loop} \, )  ,  \mu  \rangle \mapsto \langle  ( \, \ottnt{c} \, ; \, \ottkw{while} \, \ottnt{e} \, \ottkw{loop} \, \ottnt{c} \, ; \, \ottkw{end} \, \ottkw{loop} \, )  ,  \mu  \rangle }{%
{\ottdrulename{E\_While2}}{}%
}}

\newcommand{\ottdruleEXXDeclOne}[1]{\ottdrule[#1]{%
}{
 \langle  \ottkw{declare} \, \ottkw{begin} \, \ottkw{end}  ,  \mu  \rangle \mapsto \langle  \ottkw{null}  ,  \mu  \rangle }{%
{\ottdrulename{E\_Decl1}}{}%
}}

\newcommand{\ottdruleEXXDeclTwo}[1]{\ottdrule[#1]{%
\ottpremise{\langle \, \ottnt{d} \, , \, \mu \, \rangle \, \mapsto \, \langle \, \ottnt{d'} \, , \, \mu' \, \rangle}%
}{
 \langle  \ottkw{declare} \, \ottnt{d}  ,  \mu  \rangle \mapsto \langle  \ottkw{declare} \, \ottnt{d'}  ,  \mu'  \rangle }{%
{\ottdrulename{E\_Decl2}}{}%
}}

\newcommand{\ottdruleEXXForOne}[1]{\ottdrule[#1]{%
\ottpremise{ \ottnt{e}  =_ \mu   \ottnt{k} }%
\ottpremise{ \ottnt{e'}  =_ \mu   \ottnt{k'} }%
\ottpremise{ \ottnt{k}  >  \ottnt{k'} }%
}{
 \langle  ( \, \ottkw{for} \, \mathit{x} \, \ottkw{in} \, \ottnt{e} \, . \, . \, \ottnt{e'} \, \ottkw{loop} \, \ottnt{c} \, ; \, \ottkw{end} \, \ottkw{loop} \, )  ,  \mu  \rangle \mapsto \langle  \ottkw{null}  ,  \mu  \rangle }{%
{\ottdrulename{E\_For1}}{}%
}}

\newcommand{\ottdruleEXXForTwo}[1]{\ottdrule[#1]{%
\ottpremise{ \ottnt{e}  =_ \mu   \ottnt{k} }%
\ottpremise{ \ottnt{e'}  =_ \mu   \ottnt{k'} }%
\ottpremise{ \ottnt{k}  \leq  \ottnt{k'} }%
}{
 \langle  ( \, \ottkw{for} \, \mathit{x} \, \ottkw{in} \, \ottnt{e} \, . \, . \, \ottnt{e'} \, \ottkw{loop} \, \ottnt{c} \, ; \, \ottkw{end} \, \ottkw{loop} \, )  ,  \mu  \rangle \mapsto \langle  ( \, \ottkw{declare} \, \mathit{x} \, : \, \ottkw{constant} \, \ottkw{int} \, := \, \ottnt{k} \, ; \, \ottkw{begin} \, \ottnt{c} \, ; \, \ottkw{end} \, ; \, \ottkw{for} \, \mathit{x} \, \ottkw{in} \, \ottkw{\{} \, \ottnt{k} \, + \, 1 \, \ottkw{\}} \, . \, . \, \ottnt{k'} \, \ottkw{loop} \, \ottnt{c} \, ; \, \ottkw{end} \, \ottkw{loop} \, )  ,  \mu  \rangle }{%
{\ottdrulename{E\_For2}}{}%
}}

\newcommand{\ottdruleEXXProcCall}[1]{\ottdrule[#1]{%
\ottpremise{ \ottnt{e}  =_ \mu   \ottkw{proc} \, ( \, \ottcomp{\mathit{x'_{\ottmv{i}}}:\ottnt{m'_{\ottmv{i}}}\tau'_{\ottmv{i}}}{\ottmv{i}} \, ) \, \ottkw{is} \, \ottnt{d} }%
\ottpremise{( \, \ottcomp{\mathit{x'_{\ottmv{i}}}:\ottnt{m'_{\ottmv{i}}}\tau'_{\ottmv{i}}}{\ottmv{i}} \, ) \, \ottkw{\#} \, ( \, \ottcomp{\ottnt{e'_{\ottmv{j}}}}{\ottmv{j}} \, ) \, = \, [ \, \ottcomp{\mathit{x_{\ottmv{n}}}:\ottnt{m_{\ottmv{n}}}\tau_{\ottmv{n}}=\ottnt{e_{\ottmv{n}}}}{\ottmv{n}} \, ]}%
}{
 \langle  \ottnt{e} \, ( \, \ottcomp{\ottnt{e'_{\ottmv{j}}}}{\ottmv{j}} \, )  ,  \mu  \rangle \mapsto \langle  \ottkw{declare} \, [ \, \ottcomp{\mathit{x_{\ottmv{n}}}:\ottnt{m_{\ottmv{n}}}\tau_{\ottmv{n}}=\ottnt{e_{\ottmv{n}}}}{\ottmv{n}} \, ] \, \ottnt{d}  ,  \mu  \rangle }{%
{\ottdrulename{E\_ProcCall}}{}%
}}

\newcommand{\ottdefnOneStep}[1]{\begin{ottdefnblock}[#1]{$ \langle  \ottnt{c}  ,  \mu  \rangle \mapsto \langle  \ottnt{c'}  ,  \mu'  \rangle $}{\ottcom{One step evaluation}}
\ottusedrule{\ottdruleEXXNull{}}
\ottusedrule{\ottdruleEXXSeq{}}
\ottusedrule{\ottdruleEXXAssign{}}
\ottusedrule{\ottdruleEXXIfThenElseOne{}}
\ottusedrule{\ottdruleEXXIfThenElseTwo{}}
\ottusedrule{\ottdruleEXXWhileOne{}}
\ottusedrule{\ottdruleEXXWhileTwo{}}
\ottusedrule{\ottdruleEXXDeclOne{}}
\ottusedrule{\ottdruleEXXDeclTwo{}}
\ottusedrule{\ottdruleEXXForOne{}}
\ottusedrule{\ottdruleEXXForTwo{}}
\ottusedrule{\ottdruleEXXProcCall{}}
\end{ottdefnblock}}

% defn DeclEval
\newcommand{\ottdruleEXXBlockOne}[1]{\ottdrule[#1]{%
}{
\langle \, \ottkw{begin} \, \ottkw{null} \, ; \, \ottkw{end} \, , \, \mu \, \rangle \, \mapsto \, \langle \, \ottkw{begin} \, \ottkw{end} \, , \, \mu \, \rangle}{%
{\ottdrulename{E\_Block1}}{}%
}}

\newcommand{\ottdruleEXXBlockTwo}[1]{\ottdrule[#1]{%
\ottpremise{ \langle  \ottnt{c}  ,  \mu  \rangle \mapsto \langle  \ottnt{c'}  ,  \mu'  \rangle }%
}{
\langle \, \ottkw{begin} \, \ottnt{c} \, ; \, \ottkw{end} \, , \, \mu \, \rangle \, \mapsto \, \langle \, \ottkw{begin} \, \ottnt{c'} \, ; \, \ottkw{end} \, , \, \mu' \, \rangle}{%
{\ottdrulename{E\_Block2}}{}%
}}

\newcommand{\ottdruleEXXInitVarOne}[1]{\ottdrule[#1]{%
}{
\langle \, \mathit{x} \, : \, \tau \, := \, \ottnt{e} \, ; \, \ottkw{begin} \, \ottkw{end} \, , \, \mu \, \rangle \, \mapsto \, \langle \, \ottkw{begin} \, \ottkw{end} \, , \, \mu \, \rangle}{%
{\ottdrulename{E\_InitVar1}}{}%
}}

\newcommand{\ottdruleEXXInitVarTwo}[1]{\ottdrule[#1]{%
\ottpremise{ \ottnt{e}  =_ \mu   \ottnt{v} }%
\ottpremise{\langle \, \ottnt{d} \, , \, ( \, \mu \, , \, \mathit{x} \, \leftarrow \, \ottnt{v} \, ) \, \rangle \, \mapsto \, \langle \, \ottnt{d'} \, , \, ( \, \mu' \, , \, \mathit{x} \, \leftarrow \, \ottnt{v'} \, ) \, \rangle}%
}{
\langle \, \mathit{x} \, : \, \tau \, := \, \ottnt{e} \, ; \, \ottnt{d} \, , \, \mu \, \rangle \, \mapsto \, \langle \, \mathit{x} \, : \, \tau \, := \, \ottnt{v'} \, ; \, \ottnt{d'} \, , \, \mu' \, \rangle}{%
{\ottdrulename{E\_InitVar2}}{}%
}}

\newcommand{\ottdruleEXXConstOne}[1]{\ottdrule[#1]{%
}{
\langle \, \mathit{x} \, : \, \ottkw{constant} \, \tau \, := \, \ottnt{e} \, ; \, \ottkw{begin} \, \ottkw{end} \, , \, \mu \, \rangle \, \mapsto \, \langle \, \ottkw{begin} \, \ottkw{end} \, , \, \mu \, \rangle}{%
{\ottdrulename{E\_Const1}}{}%
}}

\newcommand{\ottdruleEXXConstTwo}[1]{\ottdrule[#1]{%
\ottpremise{ \ottnt{e}  =_ \mu   \ottnt{v} }%
\ottpremise{\langle \, \ottnt{d} \, [ \, \ottnt{v} \, / \, \mathit{x} \, ] \, , \, \mu \, \rangle \, \mapsto \, \langle \, \ottnt{d'} \, , \, \mu' \, \rangle}%
}{
\langle \, \mathit{x} \, : \, \ottkw{constant} \, \tau \, := \, \ottnt{e} \, ; \, \ottnt{d} \, , \, \mu \, \rangle \, \mapsto \, \langle \, \mathit{x} \, : \, \ottkw{constant} \, \tau \, := \, \ottnt{v} \, ; \, \ottnt{d'} \, , \, \mu' \, \rangle}{%
{\ottdrulename{E\_Const2}}{}%
}}

\newcommand{\ottdruleEXXProc}[1]{\ottdrule[#1]{%
}{
\langle \, \ottkw{procedure} \, \mathit{p} \, ( \, \ottcomp{\mathit{x_{\ottmv{n}}}:\ottnt{m_{\ottmv{n}}}\tau_{\ottmv{n}}}{\ottmv{n}} \, ) \, \ottkw{is} \, \ottnt{d_{{\mathrm{1}}}} \, ; \, \ottnt{d} \, , \, \mu \, \rangle \, \mapsto \, \langle \, \ottnt{d} \, [ \, \ottkw{proc} \, ( \, \ottcomp{\mathit{x_{\ottmv{n}}}:\ottnt{m_{\ottmv{n}}}\tau_{\ottmv{n}}}{\ottmv{n}} \, ) \, \ottkw{is} \, \ottnt{d_{{\mathrm{1}}}} \, / \, \mathit{p} \, ] \, , \, \mu \, \rangle}{%
{\ottdrulename{E\_Proc}}{}%
}}

\newcommand{\ottdruleEXXAliasOne}[1]{\ottdrule[#1]{%
}{
\langle \, ( \, \mathit{x} \, : \, \ottnt{m} \, \tau \, = \, \ottnt{e} \, ) \, \ottkw{begin} \, \ottkw{end} \, , \, \mu \, \rangle \, \mapsto \, \langle \, \ottkw{begin} \, \ottkw{end} \, , \, \mu \, \rangle}{%
{\ottdrulename{E\_Alias1}}{}%
}}

\newcommand{\ottdruleEXXAliasTwo}[1]{\ottdrule[#1]{%
\ottpremise{ \ottnt{e}  =_ \mu   \ottnt{v} }%
\ottpremise{\langle \, \ottnt{d} \, [ \, \ottnt{v} \, / \, \mathit{x} \, ] \, , \, \mu \, \rangle \, \mapsto \, \langle \, \ottnt{d'} \, , \, \mu' \, \rangle}%
}{
\langle \, ( \, \mathit{x} \, : \, \ottkw{in} \, \tau \, = \, \ottnt{e} \, ) \, \ottnt{d} \, , \, \mu \, \rangle \, \mapsto \, \langle \, \ottnt{d'} \, , \, \mu' \, \rangle}{%
{\ottdrulename{E\_Alias2}}{}%
}}

\newcommand{\ottdruleEXXAliasThree}[1]{\ottdrule[#1]{%
\ottpremise{\ottnt{m} \, \neq \, \ottkw{in}}%
\ottpremise{\mu \, ( \, \mathit{y} \, ) \, = \, \ottnt{v}}%
\ottpremise{\langle \, \ottnt{d} \, , \, ( \, \mu \, , \, \mathit{x} \, \leftarrow \, \ottnt{v} \, ) \, \rangle \, \mapsto \, \langle \, \ottnt{d'} \, , \, ( \, \mu' \, , \, \mathit{x} \, \leftarrow \, \ottnt{v'} \, ) \, \rangle}%
\ottpremise{\mu' \, \ottkw{\{} \, \mathit{y} \, \leftarrow \, \ottnt{v'} \, \ottkw{\}} \, \mapsto \, \mu''}%
}{
\langle \, ( \, \mathit{x} \, : \, \ottnt{m} \, \tau \, = \, \mathit{y} \, ) \, \ottnt{d} \, , \, \mu \, \rangle \, \mapsto \, \langle \, ( \, \mathit{x} \, : \, \ottnt{m} \, \tau \, = \, \mathit{y} \, ) \, \ottnt{d'} \, , \, \mu'' \, \rangle}{%
{\ottdrulename{E\_Alias3}}{}%
}}

\newcommand{\ottdruleEXXAliasesOne}[1]{\ottdrule[#1]{%
}{
\langle \, [ \, \, \, ] \, \ottnt{d} \, , \, \mu \, \rangle \, \mapsto \, \langle \, \ottnt{d} \, , \, \mu \, \rangle}{%
{\ottdrulename{E\_Aliases1}}{}%
}}

\newcommand{\ottdruleEXXAliasesTwo}[1]{\ottdrule[#1]{%
}{
\langle \, [ \, \ottcomp{\mathit{x_{\ottmv{n}}}:\ottnt{m_{\ottmv{n}}}\tau_{\ottmv{n}}=\ottnt{e_{\ottmv{n}}}}{\ottmv{n}} \, ] \, \ottkw{begin} \, \ottkw{end} \, , \, \mu \, \rangle \, \mapsto \, \langle \, \ottkw{begin} \, \ottkw{end} \, , \, \mu \, \rangle}{%
{\ottdrulename{E\_Aliases2}}{}%
}}

\newcommand{\ottdruleEXXAliasesThree}[1]{\ottdrule[#1]{%
\ottpremise{\langle \, ( \, \mathit{x} \, : \, \ottnt{m} \, \tau \, = \, \ottnt{e} \, ) \, [ \, \ottcomp{\mathit{x_{\ottmv{n}}}:\ottnt{m_{\ottmv{n}}}\tau_{\ottmv{n}}=\ottnt{e_{\ottmv{n}}}}{\ottmv{n}} \, ] \, \ottnt{d} \, , \, \mu \, \rangle \, \mapsto \, \langle \, \ottnt{d'} \, , \, \mu' \, \rangle}%
}{
\langle \, [ \, \mathit{x} \, : \, \ottnt{m} \, \tau \, = \, \ottnt{e} \, , \, \ottcomp{\mathit{x_{\ottmv{n}}}:\ottnt{m_{\ottmv{n}}}\tau_{\ottmv{n}}=\ottnt{e_{\ottmv{n}}}}{\ottmv{n}} \, ] \, \ottnt{d} \, , \, \mu \, \rangle \, \mapsto \, \langle \, \ottnt{d'} \, , \, \mu' \, \rangle}{%
{\ottdrulename{E\_Aliases3}}{}%
}}

\newcommand{\ottdefnDeclEval}[1]{\begin{ottdefnblock}[#1]{$\langle \, \ottnt{d} \, , \, \mu \, \rangle \, \mapsto \, \langle \, \ottnt{d'} \, , \, \mu' \, \rangle$}{\ottcom{Declaration evaluation}}
\ottusedrule{\ottdruleEXXBlockOne{}}
\ottusedrule{\ottdruleEXXBlockTwo{}}
\ottusedrule{\ottdruleEXXInitVarOne{}}
\ottusedrule{\ottdruleEXXInitVarTwo{}}
\ottusedrule{\ottdruleEXXConstOne{}}
\ottusedrule{\ottdruleEXXConstTwo{}}
\ottusedrule{\ottdruleEXXProc{}}
\ottusedrule{\ottdruleEXXAliasOne{}}
\ottusedrule{\ottdruleEXXAliasTwo{}}
\ottusedrule{\ottdruleEXXAliasThree{}}
\ottusedrule{\ottdruleEXXAliasesOne{}}
\ottusedrule{\ottdruleEXXAliasesTwo{}}
\ottusedrule{\ottdruleEXXAliasesThree{}}
\end{ottdefnblock}}

\newcommand{\ottdefnsevalXXcomm}{
\ottdefnStoreUpdate{}
\ottdefnManySteps{}
\ottdefnTrace{}
\ottdefnFullEvaluation{}
\ottdefnCompat{}
\ottdefnOneStep{}
\ottdefnDeclEval{}}

\newcommand{\ottdefnss}{
\ottdefnsevalXXexp
\ottdefnstyping
\ottdefnsevalXXcomm}

\newcommand{\ottall}{\ottmetavars\\[0pt]
\ottgrammar\\[5.0mm]
\ottdefnss
}

\renewenvironment{ottdefnblock}[3][]{\par\noindent \textbf{#3} \hfill \framebox{\mbox{#2}} \\[0pt]}{}
\renewcommand{\ottdrule}[4][]{\begin{array}{cc}\displaystyle\frac{#2}{#3} & (#4)\\[1em] \end{array}}
\renewcommand{\ottpremise}[1]{\hspace{0.8em} #1 \hspace{0.8em}}
\renewcommand{\ottdrulename}[1]{\textit{#1\,}}
\renewcommand{\ottkw}[1]{\mathsf{#1}}

\newcommand{\dotminus}{\frac{\cdot}{\hspace{1.5ex}}}

%\setlength\fboxrule{0pt}
%\renewcommand{\ottpremise}[1]{\framebox{\premiseSTY{#1}}}
%\renewcommand{\ottpremise}[1]{\premiseSTY{#1}}
%\renewcommand{\ottusedrule}[1]{\usedruleSTY{#1}}
%\renewcommand{\ottdrule}[4][]{\druleSTY[#1]{#2}{#3}{#4}}
%\renewenvironment{ottdefnblock}[3][]{\defnblockSTY[#1]{#2}{#3}}{\enddefnblockSTY}

%\ottstyledefaults{premiselayout=oneline}
%\ottstyledefaults{rulelayout =nobreaks}

\section*{Introduction}

We formally specified the type system and operational semantics of Loop$^{\omega}$ as described in {\cite{Crolard09}} with Ott {\cite{Sewell07}} and Isabelle/HOL proof assistant \cite{Nipkow02a}. Moreover,
both the type system and the  semantics of Loop$^{\omega}$ have been tested using Isabelle/HOL program extraction facility for inductively defined relations {\cite{Berghofer02}}. In particular, the program that computes the Ackermann function (reproduced below) type checks and behaves as expected. 

The main difference (apart from the choice of an Ada-like concrete syntax) with the description given in {\cite{Crolard09}} comes from the treatment of parameter passing. Indeed, since Ott does not currently fully support $\alpha$-conversion, we rephrased the operational semantics with explicit aliasing in order to implement the {\bf out} parameter passing mode (instead of a simpler substitution-based semantics as in  {\cite{Crolard09}}). On the other hand, the {\bf in} parameter passing mode is implemented exactly as in  {\cite{Crolard09}} and relies on Ott generated substitution (see the Isabelle/HOL code given in appendix).

Section~1 contains the description of an Ada-like grammar for Loop$^{\omega}$. We then present the type system in Section~2 and the structural operational semantic in section~3. Finally, in the appendix we include the Isabelle/HOL theory generated by Ott (all source files are available on request).

\subsubsection*{Example: the Ackermann function}

\small

\begin{lstlisting}
procedure Ack(M : in int; N : in int; R : out int) is
    P : proc(in int, out int) := Incr;
begin
    for I in 1 . . M loop
      declare
        Q : constant proc(in int, out int) := P;
        procedure Aux(S : in int; R : out int) is
          X : int := 0;
        begin
	  Q(1, X);
	  for J in 1 . . S loop
	    Q(X, X);
	  end loop;
	  R := X;
        end;
      begin
        P := Aux;
      end;
    end loop;
    P(N, R);
end;
\end{lstlisting}

\newpage

\section{Syntax}

\ottmetavars

\ottgrammar

\section{Type System}

\ottdefnstyping

\section{Structural Operational Semantics }

\ottdefnsevalXXexp
\ottdefnsevalXXcomm

\newpage

\appendix

\section{Generated Isabelle/HOL theory}

\begin{verbatim}
(* generated by Ott 0.10.17 from: _source-1-s.ott _source-1.ott _source-2-s.ott _source-2.ott 
_source-3-s.ott _source-3.ott _source-4.ott source.ott *)
theory source
imports Main Multiset
begin

(** syntax *)
types "index" = "nat"

types "ident" = "string"

types "number" = "int"

types "integer" = "int"
datatype "mode" = 
   M_In  
 | M_Out  
 | M_InOut  

types "boolean" = "bool"
datatype "ty" = 
   T_Int  
 | T_Bool  
 | T_Proc "(mode*ty) list"  
 | T_Void  

datatype "dcl" = 
   D_Empty  
 | D_Block "cmd"  
 | D_UninitVar "ident" "ty" "dcl"  
 | D_InitVar "ident" "ty" "exp" "dcl"  
 | D_Constant "ident" "ty" "exp" "dcl"  
 | D_Proc "ident" "(ident*mode*ty) list" "dcl" "dcl"  
 | D_Aliases "(ident*mode*ty*exp) list" "dcl"  
 | D_Alias "ident" "mode" "ty" "exp" "dcl"  
and "va" = 
   V_Int "integer"  
 | V_Bool "boolean"  
 | V_Proc "(ident*mode*ty) list" "dcl"  
and "cmd" = 
   C_Null  
 | C_Assign "ident" "exp"  
 | C_Seq "cmd" "cmd"  
 | C_IfThenElse "exp" "cmd" "cmd"  
 | C_While "exp" "cmd"  
 | C_Decl "dcl"  
 | C_For "ident" "exp" "exp" "cmd"  
 | C_ProcCall "exp" "exp list"  
and "exp" = 
   E_Var "ident"  
 | E_Value "va"  
 | E_Plus "exp" "exp"  
 | E_Minus "exp" "exp"  
 | E_Times "exp" "exp"  
 | E_Equal "exp" "exp"  
 | E_Greater "exp" "exp"  
 | E_Less "exp" "exp"  
 | E_And "exp" "exp"  
 | E_Or "exp" "exp"  
 | E_Not "exp"  

datatype "df" = 
   VarDecl "mode" "ty"  
 | ReturnType "ty"  

types "store" = "(ident*va) list"
types "env" = "(ident*df) list"
types "trace" = "(cmd*store) list"


(** library functions *)
lemma [mono]:"
         (!! x. f x --> g x) ==> list_all (%b. b) (map f foo_list)-->
                    list_all (%b. b) (map g foo_list) "
   apply(induct_tac foo_list, auto) done

lemma [mono]: "split f p = f (fst p) (snd p)" by (simp add: split_def)

(** subrules *)
consts
is_value_of_exp :: "exp => bool"
primrec
"is_value_of_exp (E_Var x) = (False)"
"is_value_of_exp (E_Value v) = ((True))"
"is_value_of_exp (E_Plus e1 e2) = (False)"
"is_value_of_exp (E_Minus e1 e2) = (False)"
"is_value_of_exp (E_Times e1 e2) = (False)"
"is_value_of_exp (E_Equal e1 e2) = (False)"
"is_value_of_exp (E_Greater e1 e2) = (False)"
"is_value_of_exp (E_Less e1 e2) = (False)"
"is_value_of_exp (E_And e1 e2) = (False)"
"is_value_of_exp (E_Or e1 e2) = (False)"
"is_value_of_exp (E_Not e) = (False)"


(** substitutions *)
consts
subst_ty_exp :: "exp => ident => (ty*exp) => (ty*exp)"
subst_mode_ty_exp :: "exp => ident => mode*(ty*exp) => mode*(ty*exp)"
subst_ident_mode_ty_exp :: "exp => ident => ident*(mode*ty*exp) => ident*(mode*ty*exp)"
subst_ident_mode_ty_exp_list :: "exp => ident => (ident*mode*ty*exp) list => (ident*mode*ty*exp) 
list"
subst_dcl :: "exp => ident => dcl => dcl"
subst_va :: "exp => ident => va => va"
subst_exp_list :: "exp => ident => exp list => exp list"
subst_cmd :: "exp => ident => cmd => cmd"
subst_exp :: "exp => ident => exp => exp"
primrec
"subst_ty_exp e_5 x_5 (ty1,exp1) = (ty1 , subst_exp e_5 x_5 exp1)"
"subst_mode_ty_exp e_5 x_5 (mode1,ty_exp1) = (mode1 , subst_ty_exp e_5 x_5 ty_exp1)"
"subst_ident_mode_ty_exp e_5 x_5 (ident1,mode_ty_exp1) = (ident1 , subst_mode_ty_exp e_5 x_5 
mode_ty_exp1)"
"subst_ident_mode_ty_exp_list e_5 x_5 Nil = (Nil)"
"subst_ident_mode_ty_exp_list e_5 x_5 (ident_mode_ty_exp_0#ident_mode_ty_exp_list_0) = 
((subst_ident_mode_ty_exp e_5 x_5 ident_mode_ty_exp_0) # (subst_ident_mode_ty_exp_list e_5 x_5 
ident_mode_ty_exp_list_0))"
"subst_dcl e_5 x_5 D_Empty = (D_Empty )"
"subst_dcl e_5 x_5 (D_Block c) = (D_Block (subst_cmd e_5 x_5 c))"
"subst_dcl e_5 x_5 (D_UninitVar x T d) = (D_UninitVar x T (if x_5 mem [x] then d else (subst_dcl 
e_5 x_5 d)))"
"subst_dcl e_5 x_5 (D_InitVar x T e d) = (D_InitVar x T (subst_exp e_5 x_5 e) (if x_5 mem [x] then 
d else (subst_dcl e_5 x_5 d)))"
"subst_dcl e_5 x_5 (D_Constant x T e d) = (D_Constant x T (subst_exp e_5 x_5 e) (if x_5 mem [x] 
then d else (subst_dcl e_5 x_5 d)))"
"subst_dcl e_5 x_5 (D_Proc p (x_m_T_list) d1 d2) = (D_Proc p x_m_T_list (if x_5 mem (List.map 
(%((x_0::ident),(m_0::mode),(T_0::ty)).x_0) x_m_T_list) then d1 else (subst_dcl e_5 x_5 d1)) 
(subst_dcl e_5 x_5 d2))"
"subst_dcl e_5 x_5 (D_Aliases (x_m_T_e_list) d) = (D_Aliases (subst_ident_mode_ty_exp_list e_5 x_5 
x_m_T_e_list) (if x_5 mem (List.map (%((x_0::ident),(m_0::mode),(T_0::ty),(e_0::exp)).x_0) 
x_m_T_e_list) then d else (subst_dcl e_5 x_5 d)))"
"subst_dcl e_5 x_5 (D_Alias x m T e d) = (D_Alias x m T (subst_exp e_5 x_5 e) (if x_5 mem [x] then 
d else (subst_dcl e_5 x_5 d)))"
"subst_va e5 x_5 (V_Int k) = (V_Int k)"
"subst_va e5 x_5 (V_Bool b) = (V_Bool b)"
"subst_va e5 x_5 (V_Proc (x_m_T_list) d) = (V_Proc x_m_T_list (if x_5 mem (List.map 
(%((x_0::ident),(m_0::mode),(T_0::ty)).x_0) x_m_T_list) then d else (subst_dcl e5 x_5 d)))"
"subst_exp_list e_5 x5 Nil = (Nil)"
"subst_exp_list e_5 x5 (exp_0#exp_list_0) = ((subst_exp e_5 x5 exp_0) # (subst_exp_list e_5 x5 
exp_list_0))"
"subst_cmd e_5 x5 C_Null = (C_Null )"
"subst_cmd e_5 x5 (C_Assign x e) = (C_Assign x (subst_exp e_5 x5 e))"
"subst_cmd e_5 x5 (C_Seq c1 c2) = (C_Seq (subst_cmd e_5 x5 c1) (subst_cmd e_5 x5 c2))"
"subst_cmd e_5 x5 (C_IfThenElse e c1 c2) = (C_IfThenElse (subst_exp e_5 x5 e) (subst_cmd e_5 x5 c1) 
(subst_cmd e_5 x5 c2))"
"subst_cmd e_5 x5 (C_While e c) = (C_While (subst_exp e_5 x5 e) (subst_cmd e_5 x5 c))"
"subst_cmd e_5 x5 (C_Decl d) = (C_Decl (subst_dcl e_5 x5 d))"
"subst_cmd e_5 x5 (C_For x e e' c) = (C_For x (subst_exp e_5 x5 e) (subst_exp e_5 x5 e') (if x5 mem 
[x] then c else (subst_cmd e_5 x5 c)))"
"subst_cmd e_5 x5 (C_ProcCall e (e_list)) = (C_ProcCall (subst_exp e_5 x5 e) (subst_exp_list e_5 x5 
e_list))"
"subst_exp e_5 x5 (E_Var x) = ((if x=x5 then e_5 else (E_Var x)))"
"subst_exp e_5 x5 (E_Value v) = (E_Value (subst_va e_5 x5 v))"
"subst_exp e_5 x5 (E_Plus e1 e2) = (E_Plus (subst_exp e_5 x5 e1) (subst_exp e_5 x5 e2))"
"subst_exp e_5 x5 (E_Minus e1 e2) = (E_Minus (subst_exp e_5 x5 e1) (subst_exp e_5 x5 e2))"
"subst_exp e_5 x5 (E_Times e1 e2) = (E_Times (subst_exp e_5 x5 e1) (subst_exp e_5 x5 e2))"
"subst_exp e_5 x5 (E_Equal e1 e2) = (E_Equal (subst_exp e_5 x5 e1) (subst_exp e_5 x5 e2))"
"subst_exp e_5 x5 (E_Greater e1 e2) = (E_Greater (subst_exp e_5 x5 e1) (subst_exp e_5 x5 e2))"
"subst_exp e_5 x5 (E_Less e1 e2) = (E_Less (subst_exp e_5 x5 e1) (subst_exp e_5 x5 e2))"
"subst_exp e_5 x5 (E_And e1 e2) = (E_And (subst_exp e_5 x5 e1) (subst_exp e_5 x5 e2))"
"subst_exp e_5 x5 (E_Or e1 e2) = (E_Or (subst_exp e_5 x5 e1) (subst_exp e_5 x5 e2))"
"subst_exp e_5 x5 (E_Not e) = (E_Not (subst_exp e_5 x5 e))"


(** definitions *)
(*defns eval_exp *)
inductive Fetch :: "store \<Rightarrow> ident \<Rightarrow> va \<Rightarrow> bool"
 and ExpEval :: "exp \<Rightarrow> store \<Rightarrow> va \<Rightarrow> bool"
where
(* defn Fetch *)

Fetch1I: "Fetch ( (( x , v )# mu ) ) (x) (v)"

| Fetch2I: "\<lbrakk> x  ~=  x'  ;
Fetch (mu) (x) (v)\<rbrakk> \<Longrightarrow>
Fetch ( (( x' , v' )# mu ) ) (x) (v)"

| (* defn ExpEval *)

E_ValueI: "ExpEval ((E_Value v)) (mu) (v)"

| E_IdentI: "\<lbrakk>Fetch (mu) (x) (v)\<rbrakk> \<Longrightarrow>
ExpEval ((E_Var x)) (mu) (v)"

| E_PlusI: "\<lbrakk>ExpEval (e1) (mu) ((V_Int k1)) ;
ExpEval (e2) (mu) ((V_Int k2))\<rbrakk> \<Longrightarrow>
ExpEval ((E_Plus e1 e2)) (mu) ((V_Int  ( k1  +  k2 ) ))"

| E_MinusI: "\<lbrakk>ExpEval (e1) (mu) ((V_Int k1)) ;
ExpEval (e2) (mu) ((V_Int k2))\<rbrakk> \<Longrightarrow>
ExpEval ((E_Minus e1 e2)) (mu) ((V_Int  ( k1  -  k2 ) ))"

| E_TimesI: "\<lbrakk>ExpEval (e1) (mu) ((V_Int k1)) ;
ExpEval (e2) (mu) ((V_Int k2))\<rbrakk> \<Longrightarrow>
ExpEval ((E_Times e1 e2)) (mu) ((V_Int  ( k1  *  k2 ) ))"

| E_GreaterI: "\<lbrakk>ExpEval (e1) (mu) ((V_Int k1)) ;
ExpEval (e2) (mu) ((V_Int k2))\<rbrakk> \<Longrightarrow>
ExpEval ((E_Greater e1 e2)) (mu) ((V_Bool  ( k1  >  k2 ) ))"

| E_LessI: "\<lbrakk>ExpEval (e1) (mu) ((V_Int k1)) ;
ExpEval (e2) (mu) ((V_Int k2))\<rbrakk> \<Longrightarrow>
ExpEval ((E_Less e1 e2)) (mu) ((V_Bool  ( k1  <  k2 ) ))"

| E_EqualI: "\<lbrakk>ExpEval (e1) (mu) ((V_Int k1)) ;
ExpEval (e2) (mu) ((V_Int k2))\<rbrakk> \<Longrightarrow>
ExpEval ((E_Equal e1 e2)) (mu) ((V_Bool  ( k1  =  k2 ) ))"

| E_AndI: "\<lbrakk>ExpEval (e1) (mu) ((V_Bool b1)) ;
ExpEval (e2) (mu) ((V_Bool b2))\<rbrakk> \<Longrightarrow>
ExpEval ((E_And e1 e2)) (mu) ((V_Bool  ( b1  \<and>  b2 ) ))"

| E_OrI: "\<lbrakk>ExpEval (e1) (mu) ((V_Bool b1)) ;
ExpEval (e2) (mu) ((V_Bool b2))\<rbrakk> \<Longrightarrow>
ExpEval ((E_Or e1 e2)) (mu) ((V_Bool  ( b1  \<or>  b2 ) ))"

| E_NotI: "\<lbrakk>ExpEval (e) (mu) ((V_Bool b))\<rbrakk> \<Longrightarrow>
ExpEval ((E_Not e)) (mu) ((V_Bool  (\<not>  b ) ))"

(*defns typing *)
inductive Lookup :: "ident \<Rightarrow> df \<Rightarrow> env \<Rightarrow> bool"
 and ExpTyping :: "env \<Rightarrow> exp \<Rightarrow> ty \<Rightarrow> bool"
 and LookupD :: "df \<Rightarrow> env \<Rightarrow> bool"
 and Match :: "env \<Rightarrow> exp \<Rightarrow> mode \<Rightarrow> ty \<Rightarrow> bool"
 and MatchList :: "env \<Rightarrow> exp list \<Rightarrow> (mode*ty) list \<Rightarrow> bool"
 and DeclTyping :: "env \<Rightarrow> dcl \<Rightarrow> bool"
 and CommTyping :: "env \<Rightarrow> cmd \<Rightarrow> bool"
where
(* defn Lookup *)

Lookup1I: "Lookup (x) (df) ( (( x , df ) #  G ) )"

| Lookup2I: "\<lbrakk> x  ~=  x'  ;
Lookup (x) (df) (G)\<rbrakk> \<Longrightarrow>
Lookup (x) (df) ( (( x' , df' ) #  G ) )"

| (* defn ExpTyping *)

VarI: "\<lbrakk> m  ~=  M_Out  ;
Lookup (x) ((VarDecl m T)) (G)\<rbrakk> \<Longrightarrow>
ExpTyping (G) ((E_Var x)) (T)"

| IntCstI: "ExpTyping (G) ((E_Value (V_Int  q ))) (T_Int)"

| BoolTrueI: "ExpTyping (G) ((E_Value (V_Bool  true ))) (T_Bool)"

| BoolFalseI: "ExpTyping (G) ((E_Value (V_Bool  false ))) (T_Bool)"

| PlusI: "\<lbrakk>ExpTyping (G) (e1) (T_Int) ;
ExpTyping (G) (e2) (T_Int)\<rbrakk> \<Longrightarrow>
ExpTyping (G) ((E_Plus e1 e2)) (T_Int)"

| MinusI: "\<lbrakk>ExpTyping (G) (e1) (T_Int) ;
ExpTyping (G) (e2) (T_Int)\<rbrakk> \<Longrightarrow>
ExpTyping (G) ((E_Minus e1 e2)) (T_Int)"

| TimesI: "\<lbrakk>ExpTyping (G) (e1) (T_Int) ;
ExpTyping (G) (e2) (T_Int)\<rbrakk> \<Longrightarrow>
ExpTyping (G) ((E_Times e1 e2)) (T_Int)"

| EqualI: "\<lbrakk>ExpTyping (G) (e1) (T) ;
ExpTyping (G) (e2) (T)\<rbrakk> \<Longrightarrow>
ExpTyping (G) ((E_Equal e1 e2)) (T_Bool)"

| GreaterI: "\<lbrakk>ExpTyping (G) (e1) (T_Int) ;
ExpTyping (G) (e2) (T_Int)\<rbrakk> \<Longrightarrow>
ExpTyping (G) ((E_Greater e1 e2)) (T_Bool)"

| LessI: "\<lbrakk>ExpTyping (G) (e1) (T_Int) ;
ExpTyping (G) (e2) (T_Int)\<rbrakk> \<Longrightarrow>
ExpTyping (G) ((E_Less e1 e2)) (T_Bool)"

| AndI: "\<lbrakk>ExpTyping (G) (e1) (T_Bool) ;
ExpTyping (G) (e2) (T_Bool)\<rbrakk> \<Longrightarrow>
ExpTyping (G) ((E_And e1 e2)) (T_Bool)"

| OrI: "\<lbrakk>ExpTyping (G) (e1) (T_Bool) ;
ExpTyping (G) (e2) (T_Bool)\<rbrakk> \<Longrightarrow>
ExpTyping (G) ((E_Or e1 e2)) (T_Bool)"

| NotI: "\<lbrakk>ExpTyping (G) (e) (T_Bool)\<rbrakk> \<Longrightarrow>
ExpTyping (G) ((E_Not e)) (T_Bool)"

| (* defn LookupD *)

LookupD1I: "LookupD (df) ( (( x , df ) #  G ) )"

| LookupD2I: "\<lbrakk> df  ~=  df'  ;
LookupD (df) (G)\<rbrakk> \<Longrightarrow>
LookupD (df) ( (( x , df' ) #  G ) )"

| (* defn Match *)

Match1I: "\<lbrakk>ExpTyping (G) (e) (T)\<rbrakk> \<Longrightarrow>
Match (G) (e) (M_In) (T)"

| Match2I: "\<lbrakk>Lookup (x) ((VarDecl m T)) (G) ;
 m  ~=  M_In \<rbrakk> \<Longrightarrow>
Match (G) ((E_Var x)) (M_Out) (T)"

| Match3I: "\<lbrakk>Lookup (x) ((VarDecl M_InOut T)) (G)\<rbrakk> \<Longrightarrow>
Match (G) ((E_Var x)) (M_InOut) (T)"

| (* defn MatchList *)

MatchList1I: "MatchList (G) ([]) ([])"

| MatchList2I: "\<lbrakk>Match (G) (e) (m) (T) ;
MatchList (G) ((e_list)) ((m_T_list))\<rbrakk> \<Longrightarrow>
MatchList (G) (((e) # e_list)) (((m,T) # m_T_list))"

| (* defn DeclTyping *)

EmptyI: "DeclTyping (G) (D_Empty)"

| BlockI: "\<lbrakk>CommTyping (G) (c)\<rbrakk> \<Longrightarrow>
DeclTyping (G) ((D_Block c))"

| UninitVarI: "\<lbrakk>DeclTyping ( (( x , (VarDecl M_InOut T) ) #  G ) ) (d)\<rbrakk> 
\<Longrightarrow>
DeclTyping (G) ((D_UninitVar x T d))"

| InitVarI: "\<lbrakk>ExpTyping (G) (e) (T) ;
DeclTyping ( (( x , (VarDecl M_InOut T) ) #  G ) ) (d)\<rbrakk> \<Longrightarrow>
DeclTyping (G) ((D_InitVar x T e d))"

| ConstantI: "\<lbrakk>ExpTyping (G) (e) (T) ;
DeclTyping ( (( x , (VarDecl M_In T) ) #  G ) ) (d)\<rbrakk> \<Longrightarrow>
DeclTyping (G) ((D_Constant x T e d))"

| ProcI: "\<lbrakk>DeclTyping ( (   (List.rev  ((List.map 
(%((x_0::ident),(m_0::mode),(T_0::ty)).(x_0,(VarDecl m_0 T_0))) x_m_T_list))  @ G)   ) ) (d1) ;
DeclTyping ( (( p , (VarDecl M_In (T_Proc ((List.map 
(%((x_0::ident),(m_0::mode),(T_0::ty)).(m_0,T_0)) x_m_T_list)))) ) #  G ) ) (d2)\<rbrakk> 
\<Longrightarrow>
DeclTyping (G) ((D_Proc p (x_m_T_list) d1 d2))"

| (* defn CommTyping *)

NullI: "CommTyping (G) (C_Null)"

| SeqI: "\<lbrakk>CommTyping (G) (c1) ;
CommTyping (G) (c2)\<rbrakk> \<Longrightarrow>
CommTyping (G) ((C_Seq c1 c2))"

| AssignI: "\<lbrakk> m  ~=  M_In  ;
Lookup (x) ((VarDecl m T)) (G) ;
ExpTyping (G) (e) (T)\<rbrakk> \<Longrightarrow>
CommTyping (G) ((C_Assign x e))"

| IfThenElseI: "\<lbrakk>ExpTyping (G) (e) (T_Bool) ;
CommTyping (G) (c1) ;
CommTyping (G) (c2)\<rbrakk> \<Longrightarrow>
CommTyping (G) ((C_IfThenElse e c1 c2))"

| WhileI: "\<lbrakk>ExpTyping (G) (e) (T_Bool) ;
CommTyping (G) (c)\<rbrakk> \<Longrightarrow>
CommTyping (G) ((C_While e c))"

| ForI: "\<lbrakk>ExpTyping (G) (e) (T_Int) ;
ExpTyping (G) (e') (T_Int) ;
CommTyping ( (( x , (VarDecl M_In T_Int) ) #  G ) ) (c)\<rbrakk> \<Longrightarrow>
CommTyping (G) ((C_For x e e' c))"

| DeclI: "\<lbrakk>DeclTyping (G) (d)\<rbrakk> \<Longrightarrow>
CommTyping (G) ((C_Decl d))"

| ProcCallI: "\<lbrakk>ExpTyping (G) (e) ((T_Proc (m_T_list))) ;
MatchList (G) ((e_list)) ((m_T_list))\<rbrakk> \<Longrightarrow>
CommTyping (G) ((C_ProcCall e (e_list)))"

(*defns eval_comm *)
inductive StoreUpdate :: "store \<Rightarrow> ident \<Rightarrow> va \<Rightarrow> store 
\<Rightarrow> bool"
 and ManySteps :: "cmd \<Rightarrow> store \<Rightarrow> integer \<Rightarrow> cmd \<Rightarrow> 
store \<Rightarrow> bool"
 and Trace :: "cmd \<Rightarrow> store \<Rightarrow> integer \<Rightarrow> trace \<Rightarrow> bool"
 and FullEvaluation :: "cmd \<Rightarrow> store \<Rightarrow> store \<Rightarrow> bool"
 and Compat :: "(ident*mode*ty) list \<Rightarrow> exp list \<Rightarrow> (ident*mode*ty*exp) list 
\<Rightarrow> bool"
 and OneStep :: "cmd \<Rightarrow> store \<Rightarrow> cmd \<Rightarrow> store \<Rightarrow> bool"
 and DeclEval :: "dcl \<Rightarrow> store \<Rightarrow> dcl \<Rightarrow> store \<Rightarrow> bool"
where
(* defn StoreUpdate *)

Update1I: "StoreUpdate ( (( x , v )# mu ) ) (x) (v') ( (( x , v' )# mu ) )"

| Update2I: "\<lbrakk> x  ~=  x'  ;
StoreUpdate (mu) (x) (v') (mu')\<rbrakk> \<Longrightarrow>
StoreUpdate ( (( x' , v )# mu ) ) (x) (v') ( (( x' , v )# mu' ) )"

| (* defn ManySteps *)

ManySteps1I: "ManySteps (c) (mu) ( 0 ) (c) (mu)"

| ManySteps2I: "ManySteps (C_Null) (mu) (k) (C_Null) (mu)"

| ManySteps3I: "\<lbrakk>OneStep (c) (mu) (c') (mu') ;
ManySteps (c') (mu') ( ( k  -   1  ) ) (c'') (mu'')\<rbrakk> \<Longrightarrow>
ManySteps (c) (mu) (k) (c'') (mu'')"

| (* defn Trace *)

Trace1I: "Trace (c) (mu) ( 0 ) ( [] )"

| Trace2I: "Trace (C_Null) (mu) (k) ( [] )"

| Trace3I: "\<lbrakk>OneStep (c) (mu) (c') (mu') ;
Trace (c') (mu') ( ( k  -   1  ) ) ( (c'_mu'_list) )\<rbrakk> \<Longrightarrow>
Trace (c) (mu) (k) ( ((c',mu') # c'_mu'_list) )"

| (* defn FullEvaluation *)

Eval1I: "FullEvaluation (C_Null) (mu) (mu)"

| Eval2I: "\<lbrakk>OneStep (c) (mu) (c') (mu') ;
FullEvaluation (c') (mu') (mu'')\<rbrakk> \<Longrightarrow>
FullEvaluation (c) (mu) (mu'')"

| (* defn Compat *)

E_Compat1I: "Compat ([]) ([]) ([])"

| E_Compat2I: "\<lbrakk>Compat ((x'_m'_T'_list)) ((e'_list)) ((x_m_T_e_list))\<rbrakk> 
\<Longrightarrow>
Compat (((x,m,T) # x'_m'_T'_list)) (((e) # e'_list)) (((x,m,T,e) # x_m_T_e_list))"

| (* defn OneStep *)

E_NullI: "OneStep ( (C_Seq C_Null c) ) (mu) (c) (mu)"

| E_SeqI: "\<lbrakk>OneStep (c1) (mu) (c1') (mu')\<rbrakk> \<Longrightarrow>
OneStep ( (C_Seq c1 c2) ) (mu) ( (C_Seq c1' c2) ) (mu')"

| E_AssignI: "\<lbrakk>ExpEval (e) (mu) (v) ;
StoreUpdate (mu) (x) (v) (mu')\<rbrakk> \<Longrightarrow>
OneStep ( (C_Assign x e) ) (mu) (C_Null) (mu')"

| E_IfThenElse1I: "\<lbrakk>ExpEval (e) (mu) ((V_Bool  true ))\<rbrakk> \<Longrightarrow>
OneStep ( (C_IfThenElse e c1 c2) ) (mu) (c1) (mu)"

| E_IfThenElse2I: "\<lbrakk>ExpEval (e) (mu) ((V_Bool  false ))\<rbrakk> \<Longrightarrow>
OneStep ( (C_IfThenElse e c1 c2) ) (mu) (c2) (mu)"

| E_While1I: "\<lbrakk>ExpEval (e) (mu) ((V_Bool  false ))\<rbrakk> \<Longrightarrow>
OneStep ( (C_While e c) ) (mu) (C_Null) (mu)"

| E_While2I: "\<lbrakk>ExpEval (e) (mu) ((V_Bool  true ))\<rbrakk> \<Longrightarrow>
OneStep ( (C_While e c) ) (mu) ( (C_Seq c (C_While e c)) ) (mu)"

| E_Decl1I: "OneStep ((C_Decl D_Empty)) (mu) (C_Null) (mu)"

| E_Decl2I: "\<lbrakk>DeclEval (d) (mu) (d') (mu')\<rbrakk> \<Longrightarrow>
OneStep ((C_Decl d)) (mu) ((C_Decl d')) (mu')"

| E_For1I: "\<lbrakk>ExpEval (e) (mu) ((V_Int k)) ;
ExpEval (e') (mu) ((V_Int k')) ;
 ( k  >  k' ) \<rbrakk> \<Longrightarrow>
OneStep ( (C_For x e e' c) ) (mu) (C_Null) (mu)"

| E_For2I: "\<lbrakk>ExpEval (e) (mu) ((V_Int k)) ;
ExpEval (e') (mu) ((V_Int k')) ;
 ( k  <=  k' ) \<rbrakk> \<Longrightarrow>
OneStep ( (C_For x e e' c) ) (mu) ( (C_Seq (C_Decl (D_Constant x T_Int (E_Value (V_Int k)) (D_Block 
c))) (C_For x (E_Value (V_Int  ( k  +   1  ) )) (E_Value (V_Int k')) c)) ) (mu)"

| E_ProcCallI: "\<lbrakk>ExpEval (e) (mu) ((V_Proc (x'_m'_T'_list) d)) ;
Compat ((x'_m'_T'_list)) ((e'_list)) ((x_m_T_e_list))\<rbrakk> \<Longrightarrow>
OneStep ((C_ProcCall e (e'_list))) (mu) ((C_Decl (D_Aliases (x_m_T_e_list) d))) (mu)"

| (* defn DeclEval *)

E_Block1I: "DeclEval ((D_Block C_Null)) (mu) (D_Empty) (mu)"

| E_Block2I: "\<lbrakk>OneStep (c) (mu) (c') (mu')\<rbrakk> \<Longrightarrow>
DeclEval ((D_Block c)) (mu) ((D_Block c')) (mu')"

| E_InitVar1I: "DeclEval ((D_InitVar x T e D_Empty)) (mu) (D_Empty) (mu)"

| E_InitVar2I: "\<lbrakk>ExpEval (e) (mu) (v) ;
DeclEval (d) ( (( x , v )# mu ) ) (d') ( (( x , v' )# mu' ) )\<rbrakk> \<Longrightarrow>
DeclEval ((D_InitVar x T e d)) (mu) ((D_InitVar x T (E_Value v') d')) (mu')"

| E_Const1I: "DeclEval ((D_Constant x T e D_Empty)) (mu) (D_Empty) (mu)"

| E_Const2I: "\<lbrakk>ExpEval (e) (mu) (v) ;
DeclEval ( (subst_dcl (E_Value  v )  x   d ) ) (mu) (d') (mu')\<rbrakk> \<Longrightarrow>
DeclEval ((D_Constant x T e d)) (mu) ((D_Constant x T (E_Value v) d')) (mu')"

| E_ProcI: "DeclEval ((D_Proc p (x_m_T_list) d1 d)) (mu) ( (subst_dcl (E_Value  (V_Proc 
(x_m_T_list) d1) )  p   d ) ) (mu)"

| E_Alias1I: "DeclEval ((D_Alias x m T e D_Empty)) (mu) (D_Empty) (mu)"

| E_Alias2I: "\<lbrakk>ExpEval (e) (mu) (v) ;
DeclEval ( (subst_dcl (E_Value  v )  x   d ) ) (mu) (d') (mu')\<rbrakk> \<Longrightarrow>
DeclEval ((D_Alias x M_In T e d)) (mu) (d') (mu')"

| E_Alias3I: "\<lbrakk> m  ~=  M_In  ;
Fetch (mu) (y) (v) ;
DeclEval (d) ( (( x , v )# mu ) ) (d') ( (( x , v' )# mu' ) ) ;
StoreUpdate (mu') (y) (v') (mu'')\<rbrakk> \<Longrightarrow>
DeclEval ((D_Alias x m T (E_Var y) d)) (mu) ((D_Alias x m T (E_Var y) d')) (mu'')"

| E_Aliases1I: "DeclEval ((D_Aliases [] d)) (mu) (d) (mu)"

| E_Aliases2I: "DeclEval ((D_Aliases (x_m_T_e_list) D_Empty)) (mu) (D_Empty) (mu)"

| E_Aliases3I: "\<lbrakk>DeclEval ((D_Alias x m T e (D_Aliases (x_m_T_e_list) d))) (mu) (d') 
(mu')\<rbrakk> \<Longrightarrow>
DeclEval ((D_Aliases ((x,m,T,e) # x_m_T_e_list) d)) (mu) (d') (mu')"


code_module Evaluation
contains

test1 = "ExpEval ( ( (E_Plus (E_Value (V_Int  2 )) (E_Value (V_Int  3 ))) ) ) ( Nil ) ( _ )"
test2 = "ExpEval ( ( (E_Plus (E_Var ''X'') (E_Value (V_Int  3 ))) ) ) ( ([(''X'',(V_Int  5 ))]) ) ( 
_ )"

ML {* DSeq.hd Evaluation.test1 *}


code_module Typing (* file "Typing.sml" *)
contains

test1 = "ExpTyping ( Nil ) ((E_Var ''X'')) (T_Int)"

test2 = "ExpTyping ( ((''X'',(VarDecl  M_In  T_Int)) # [(''Y'',(VarDecl  M_In  T_Int))]) ) 
((E_Equal  ( (E_Plus (E_Var ''X'') (E_Value (V_Int  1 ))) )  (E_Var ''Y''))) (T_Bool)"

ML {* Typing.test1 *}
ML {* Typing.test2 *}


code_module Evaluation
contains

test1 = "ExpEval ( ( (E_Plus (E_Value (V_Int  2 )) (E_Value (V_Int  3 ))) ) ) ( Nil ) ( _ )"

test2 = "StoreUpdate ( ((''X'',(V_Int  2 )) # [(''Y'',(V_Int  3 ))]) ) (''X'') ((V_Int  3 )) ( _ )"

test3 = "ManySteps ( (C_Assign ''X'' (E_Plus (E_Var ''X'') (E_Value (V_Int  1 )))) ) ( 
([(''X'',(V_Int  2 ))]) ) ( 1 ) ( _ ) ( _ )"

test4 = "FullEvaluation ( (C_Seq (C_Assign ''X'' (E_Plus (E_Var ''X'') (E_Var ''Y''))) (C_Assign 
''Y'' (E_Plus (E_Var ''X'') (E_Var ''Y'')))) ) ( ((''X'',(V_Int  42 )) # [(''Y'',(V_Int  12 ))]) ) 
( _ )"

test5 = "FullEvaluation ( (C_IfThenElse (E_Var ''B'') (C_Assign ''X'' (E_Value (V_Int  1 ))) 
(C_Assign ''Y'' (E_Value (V_Int  1 )))) ) ( ((''B'',(V_Bool  true )) # (''X'',(V_Int  0 )) # 
[(''Y'',(V_Int  0 ))]) ) ( _ )"

ML {* DSeq.hd Evaluation.test1 *}
ML {* DSeq.hd Evaluation.test2 *}
ML {* DSeq.hd Evaluation.test3 *}
ML {* DSeq.hd Evaluation.test4 *}
ML {* DSeq.hd Evaluation.test5 *}


code_module Typing (* file "Extraction.sml" *)
contains

test1 = "CommTyping ( ([(''X'',(VarDecl M_InOut T_Int))]) ) ((C_Assign ''X'' (E_Plus (E_Var ''X'') 
(E_Value (V_Int  1 )))))"

test2 = "CommTyping ( ((''X'',(VarDecl M_InOut T_Int)) # (''Y'',(VarDecl M_In T_Bool)) # 
[(''B'',(VarDecl M_In T_Bool))]) ) ((C_IfThenElse (E_Var ''B'') (C_Assign ''X'' (E_Value (V_Int  1 
))) (C_Assign ''Y'' (E_Value (V_Int  1 )))))"

ML {* Typing.test1 *}
ML {* Typing.test2 *}


code_module Evaluation
contains

test1 = "FullEvaluation ((C_Decl (D_Constant ''B'' T_Bool (E_Value (V_Bool  false )) (D_Block 
(C_IfThenElse (E_Var ''B'') (C_Assign ''X'' (E_Value (V_Int  1 ))) (C_Assign ''Y'' (E_Value (V_Int  
1 )))))))) ( ((''X'',(V_Int  0 )) # [(''Y'',(V_Int  0 ))]) ) ( _ )"

test2 = "FullEvaluation ((C_For ''I'' (E_Value (V_Int  1 )) (E_Var ''X'') (C_Assign ''Y'' (E_Plus 
(E_Var ''Y'') (E_Var ''X''))))) ( ((''X'',(V_Int  5 )) # [(''Y'',(V_Int  0 ))]) ) ( _ )"

ML {* DSeq.hd Evaluation.test1 *}
ML {* DSeq.hd Evaluation.test2 *}


code_module Typing (* file "Extraction.sml" *)
contains

test1 = "CommTyping ( ([(''X'',(VarDecl M_InOut T_Int))]) ) ((C_Assign ''X'' (E_Plus (E_Var ''X'') 
(E_Value (V_Int  1 )))))"

test2 = "CommTyping ( ([(''X'',(VarDecl M_InOut T_Int))]) ) ((C_Decl (D_InitVar ''X'' T_Int 
(E_Value (V_Int  42 )) (D_Block (C_Seq (C_Assign ''X'' (E_Plus (E_Var ''Y'') (E_Value (V_Int  1 
)))) (C_Seq (C_Assign ''X'' (E_Plus (E_Var ''X'') (E_Value (V_Int  1 )))) (C_Assign ''Y'' (E_Minus 
(E_Var ''Y'') (E_Value (V_Int  1 ))))))))))"

test3 = "CommTyping ( ([(''X'',(VarDecl M_InOut T_Int))]) ) ((C_Decl (D_InitVar ''Y'' T_Bool 
(E_Value (V_Bool  false )) (D_Block (C_For ''I'' (E_Value (V_Int  1 )) (E_Var ''X'') (C_Assign 
''X'' (E_Plus (E_Var ''Y'') (E_Value (V_Int  1 )))))))))"

ML {* Typing.test1 *}
ML {* Typing.test2 *}
ML {* Typing.test3 *}


code_module Typing
contains

test1 = "CommTyping ( ([(''R'',(VarDecl M_Out T_Bool))]) ) ((C_Decl (D_InitVar ''Y'' T_Int (E_Value 
(V_Int  42 )) (D_Proc ''P'' ((''I'',M_InOut,T_Int) # [(''B'',M_Out,T_Bool)]) (D_Block (C_Assign 
''B''  ( (E_Equal (E_Var ''I'') (E_Value (V_Int  1 ))) ) )) (D_Block (C_ProcCall (E_Var ''P'') 
(((E_Var ''Y'')) # [((E_Var ''R''))])))))))"

test2 = "CommTyping ( ([(''R'',(VarDecl M_Out T_Int))]) ) ((C_Decl (D_Proc ''Incr'' 
((''N'',M_In,T_Int) # [(''R'',M_Out,T_Int)]) (D_Block (C_Assign ''R'' (E_Plus (E_Var ''N'') 
(E_Value (V_Int  1 ))))) (D_Proc ''Ack'' ((''M'',M_In,T_Int) # (''N'',M_In,T_Int) # 
[(''R'',M_Out,T_Int)]) (D_InitVar ''P'' (T_Proc ((M_In,T_Int) # [(M_Out,T_Int)])) (E_Var ''Incr'') 
(D_Block (C_Seq (C_For ''I'' (E_Value (V_Int  1 )) (E_Var ''M'') (C_Decl (D_Proc ''Aux'' 
((''S'',M_In,T_Int) # [(''R'',M_Out,T_Int)]) (D_InitVar ''X'' T_Int (E_Value (V_Int  0 )) (D_Block 
(C_Seq (C_ProcCall (E_Var ''P'') (((E_Value (V_Int  1 ))) # [((E_Var ''X''))])) (C_Seq (C_For ''J'' 
(E_Value (V_Int  1 )) (E_Var ''S'') (C_ProcCall (E_Var ''P'') (((E_Var ''X'')) # [((E_Var 
''X''))]))) (C_Assign ''R'' (E_Var ''X'')))))) (D_Block (C_Assign ''P'' (E_Var ''Aux'')))))) 
(C_ProcCall (E_Var ''P'') (((E_Var ''N'')) # [((E_Var ''R''))]))))) (D_Block (C_ProcCall (E_Var 
''Ack'') (((E_Value (V_Int  2 ))) # ((E_Value (V_Int  2 ))) # [((E_Var ''R''))])))))))"

test24 = "CommTyping ( ([(''R'',(VarDecl M_Out T_Int))]) ) ((C_Decl (D_Proc ''Comp'' 
([(''P1'',M_In,(T_Proc ((M_In,T_Int) # [(M_Out,T_Int)])))] @ [(''P2'',M_In,(T_Proc ((M_In,T_Int) # 
[(M_Out,T_Int)])))] @ [(''P3'',M_Out,(T_Proc ((M_In,T_Int) # [(M_Out,T_Int)])))]) (D_Proc ''P'' 
((''N'',M_In,T_Int) # [(''R'',M_Out,T_Int)]) (D_InitVar ''X'' T_Int (E_Value (V_Int  0 )) (D_Block 
(C_Seq (C_ProcCall (E_Var ''P1'') (((E_Var ''N'')) # [((E_Var ''X''))])) (C_ProcCall (E_Var ''P2'') 
(((E_Var ''X'')) # [((E_Var ''R''))]))))) (D_Block (C_Assign ''P3'' (E_Var ''P'')))) (D_Proc 
''Incr'' ((''N'',M_In,T_Int) # [(''R'',M_Out,T_Int)]) (D_Block (C_Assign ''R'' (E_Plus (E_Var 
''N'') (E_Value (V_Int  1 ))))) (D_Proc ''IncrN'' ((''M'',M_In,T_Int) # (''N'',M_In,T_Int) # 
[(''R'',M_Out,T_Int)]) (D_InitVar ''P'' (T_Proc ((M_In,T_Int) # [(M_Out,T_Int)])) (E_Var ''Incr'') 
(D_Block (C_Seq (C_For ''I'' (E_Value (V_Int  1 )) (E_Var ''N'') (C_ProcCall (E_Var ''Comp'') 
(((E_Var ''P'')) # ((E_Var ''P'')) # [((E_Var ''P''))]))) (C_ProcCall (E_Var ''P'') (((E_Var 
''M'')) # [((E_Var ''R''))]))))) (D_Block (C_ProcCall (E_Var ''IncrN'') (((E_Value (V_Int  3 ))) # 
((E_Value (V_Int  3 ))) # [((E_Var ''R''))]))))))))"

ML {* Typing.test1 *}
ML {* Typing.test2 *}
ML {* Typing.test24 *}


code_module Evaluation
contains

test1 = "ManySteps ((C_Decl (D_InitVar ''Y'' T_Int (E_Value (V_Int  42 )) (D_Proc ''P'' 
((''I'',M_InOut,T_Int) # [(''B'',M_Out,T_Bool)]) (D_Block (C_Assign ''B''  ( (E_Equal (E_Var ''I'') 
(E_Value (V_Int  1 ))) ) )) (D_Block (C_ProcCall (E_Var ''P'') (((E_Var ''Y'')) # [((E_Var 
''R''))]))))))) ( ([(''R'',(V_Bool  false ))]) ) ( 1 ) ( _ ) ( _ )"

test2 = "ManySteps ((C_Decl (D_InitVar ''Y'' T_Int (E_Value (V_Int  42 )) (D_Proc ''P'' 
((''I'',M_InOut,T_Int) # [(''B'',M_Out,T_Bool)]) (D_Block (C_Assign ''B''  ( (E_Equal (E_Var ''I'') 
(E_Value (V_Int  1 ))) ) )) (D_Block (C_ProcCall (E_Var ''P'') (((E_Var ''Y'')) # [((E_Var 
''R''))]))))))) ( ([(''R'',(V_Bool  false ))]) ) ( 2 ) ( _ ) ( _ )"
  
test3 = "ManySteps ((C_Decl (D_InitVar ''Y'' T_Int (E_Value (V_Int  42 )) (D_Proc ''P'' 
((''I'',M_In,T_Int) # [(''B'',M_Out,T_Bool)]) (D_Block (C_Assign ''B''  ( (E_Equal (E_Var ''I'') 
(E_Value (V_Int  1 ))) ) )) (D_Block (C_ProcCall (E_Var ''P'') (((E_Var ''Y'')) # [((E_Var 
''R''))]))))))) ( ([(''R'',(V_Bool  false ))]) ) ( 3 ) ( _ ) ( _ )"

test10 = "FullEvaluation ((C_Decl (D_InitVar ''Y'' T_Int (E_Value (V_Int  42 )) (D_Proc ''P'' 
((''I'',M_In,T_Int) # [(''B'',M_Out,T_Bool)]) (D_Block (C_Assign ''B''  ( (E_Equal (E_Var ''I'') 
(E_Value (V_Int  1 ))) ) )) (D_Block (C_ProcCall (E_Var ''P'') (((E_Var ''Y'')) # [((E_Var 
''R''))]))))))) ( ([(''R'',(V_Bool  true ))]) ) ( _ )"

test20 = "Trace ((C_Decl (D_InitVar ''Y'' T_Int (E_Value (V_Int  42 )) (D_Proc ''P'' 
((''I'',M_In,T_Int) # [(''B'',M_Out,T_Bool)]) (D_Block (C_Assign ''B''  ( (E_Equal (E_Var ''I'') 
(E_Value (V_Int  1 ))) ) )) (D_Block (C_ProcCall (E_Var ''P'') (((E_Var ''Y'')) # [((E_Var 
''R''))]))))))) ( ([(''R'',(V_Bool  false ))]) ) ( 20 ) ( _ )"

test24 = "FullEvaluation ((C_Decl (D_Proc ''Incr'' ((''N'',M_In,T_Int) # [(''R'',M_Out,T_Int)]) 
(D_Block (C_Assign ''R'' (E_Plus (E_Var ''N'') (E_Value (V_Int  1 ))))) (D_Proc ''Plus'' 
((''M'',M_In,T_Int) # (''N'',M_In,T_Int) # [(''R'',M_Out,T_Int)]) (D_InitVar ''X'' T_Int (E_Var 
''M'') (D_Block (C_Seq (C_For ''I'' (E_Value (V_Int  1 )) (E_Var ''N'') (C_ProcCall (E_Var 
''Incr'') (((E_Var ''X'')) # [((E_Var ''X''))]))) (C_Assign ''R'' (E_Var ''X''))))) (D_Block 
(C_ProcCall (E_Var ''Plus'') (((E_Value (V_Int  3 ))) # ((E_Value (V_Int  5 ))) # [((E_Var 
''R''))]))))))) ( ([(''R'',(V_Int  0 ))]) ) ( _ )"

test25 = "FullEvaluation ((C_Decl (D_Proc ''Comp'' ([(''P1'',M_In,(T_Proc ((M_In,T_Int) # 
[(M_Out,T_Int)])))] @ [(''P2'',M_In,(T_Proc ((M_In,T_Int) # [(M_Out,T_Int)])))] @ 
[(''P3'',M_Out,(T_Proc ((M_In,T_Int) # [(M_Out,T_Int)])))]) (D_Proc ''P'' ((''N'',M_In,T_Int) # 
[(''R'',M_Out,T_Int)]) (D_InitVar ''X'' T_Int (E_Value (V_Int  0 )) (D_Block (C_Seq (C_ProcCall 
(E_Var ''P1'') (((E_Var ''N'')) # [((E_Var ''X''))])) (C_ProcCall (E_Var ''P2'') (((E_Var ''X'')) # 
[((E_Var ''R''))]))))) (D_Block (C_Assign ''P3'' (E_Var ''P'')))) (D_Proc ''Incr'' 
((''N'',M_In,T_Int) # [(''R'',M_Out,T_Int)]) (D_Block (C_Assign ''R'' (E_Plus (E_Var ''N'') 
(E_Value (V_Int  1 ))))) (D_Proc ''IncrN'' ((''M'',M_In,T_Int) # (''N'',M_In,T_Int) # 
[(''R'',M_Out,T_Int)]) (D_InitVar ''P'' (T_Proc ((M_In,T_Int) # [(M_Out,T_Int)])) (E_Var ''Incr'') 
(D_Block (C_Seq (C_For ''I'' (E_Value (V_Int  1 )) (E_Var ''N'') (C_ProcCall (E_Var ''Comp'') 
(((E_Var ''P'')) # ((E_Var ''P'')) # [((E_Var ''P''))]))) (C_ProcCall (E_Var ''P'') (((E_Var 
''M'')) # [((E_Var ''R''))]))))) (D_Block (C_ProcCall (E_Var ''IncrN'') (((E_Value (V_Int  3 ))) # 
((E_Value (V_Int  3 ))) # [((E_Var ''R''))])))))))) ( ([(''R'',(V_Int  0 ))]) ) ( _ )"

test30 = "FullEvaluation ((C_Decl (D_Proc ''Incr'' ((''N'',M_In,T_Int) # [(''R'',M_Out,T_Int)]) 
(D_Block (C_Assign ''R'' (E_Plus (E_Var ''N'') (E_Value (V_Int  1 ))))) (D_Proc ''Ack'' 
((''M'',M_In,T_Int) # (''N'',M_In,T_Int) # [(''R'',M_Out,T_Int)]) (D_InitVar ''P'' (T_Proc 
((M_In,T_Int) # [(M_Out,T_Int)])) (E_Var ''Incr'') (D_Block (C_Seq (C_For ''I'' (E_Value (V_Int  1 
)) (E_Var ''M'') (C_Decl (D_Constant ''Q'' (T_Proc ((M_In,T_Int) # [(M_Out,T_Int)])) (E_Var ''P'') 
(D_Proc ''Aux'' ((''S'',M_In,T_Int) # [(''R'',M_Out,T_Int)]) (D_InitVar ''X'' T_Int (E_Value (V_Int 
 0 )) (D_Block (C_Seq (C_ProcCall (E_Var ''Q'') (((E_Value (V_Int  1 ))) # [((E_Var ''X''))])) 
(C_Seq (C_For ''J'' (E_Value (V_Int  1 )) (E_Var ''S'') (C_ProcCall (E_Var ''Q'') (((E_Var ''X'')) 
# [((E_Var ''X''))]))) (C_Assign ''R'' (E_Var ''X'')))))) (D_Block (C_Assign ''P'' (E_Var 
''Aux''))))))) (C_ProcCall (E_Var ''P'') (((E_Var ''N'')) # [((E_Var ''R''))]))))) (D_Block 
(C_ProcCall (E_Var ''Ack'') (((E_Value (V_Int  3 ))) # ((E_Value (V_Int  2 ))) # [((E_Var 
''R''))]))))))) ( ([(''R'',(V_Int  0 ))]) ) ( _ )"

ML {* print_depth 1000 *}
ML {* DSeq.hd Evaluation.test24 *}
ML {* DSeq.hd Evaluation.test25 *}
ML {* DSeq.hd Evaluation.test30 *}
ML {* val trace = DSeq.hd Evaluation.test24 *}
ML {* List.nth (trace, 0) *}
ML {* List.nth (trace, 1) *}
ML {* List.nth (trace, 2) *}




end
\end{verbatim}

\newpage

\bibliography{biblio}
\bibliographystyle{alpha}

\end{document}